\documentclass[iop,twocolappendix]{emulateapj}
\usepackage[utf8]{inputenc}
\usepackage{amsmath}
\usepackage{amssymb}
\usepackage{verbatim}
\usepackage{natbib}
\usepackage{graphicx,subfigure}   
\usepackage[colorlinks,urlcolor=blue,citecolor=blue,linkcolor=blue]{hyperref}
\renewcommand{\b}[1]{\boldsymbol{#1}} 
\newcommand{\ud}{{\rm d}} 

\newcommand{\cross}{\times}
\newcommand{\bmath}[1]{\mbox{\boldmath{$#1$}}}

\newcommand{\bi}{\begin{itemize}}
\newcommand{\ei}{\end{itemize}}

\newcommand{\parr}{\parallel}

\def\compe{\,c/\omega_{pe}}
\newcommand{\rli}{\,r_{\rm Li}}

\newcommand{\eq}[1]{Equation \eqref{eq:#1}}
\newcommand{\xsh}{x-x_{\rm sh}}
\newcommand{\xshnorm}{(\xsh)/\rli}

\newcommand{\fig}[1]{Figure~\ref{fig:#1}}

\newcommand{\sect}[1]{Section \ref{sec:#1}}
\newcommand{\app}[1]{Appendix \ref{sec:#1}}

\newcommand{\be}{\begin{eqnarray}}
\newcommand{\ee}{\end{eqnarray}}
\newcommand{\qt}{(1+q\,t)}

\newcommand{\dby}{\delta B_y}
\newcommand{\dbx}{\delta B_x}
\newcommand{\dbz}{\delta B_z}

\newcommand{\perpe}{{e,\perp}}
\newcommand{\pare}{{e,\parallel}}
\newcommand{\perpi}{{i,\perp}}
\newcommand{\pari}{{i,\parallel}}
\newcommand{\alf}{Alfv\'en}

\def\L{\bmath{{L}}}
\def\bvec{\bmath{B}}

\def\evec{\bmath{E}}

\newcommand{\Teperp}{T_{e,\perp} }

\begin{document}
\title{Electron Heating in Low Mach Number Perpendicular shocks. I. Heating Mechanism}
\author{Xinyi Guo,$^1$ Lorenzo Sironi,$^2$ and Ramesh Narayan$^1$}
\affil{$^1$Harvard-Smithsonian Center for Astrophysics, 
60 Garden Street, Cambridge, MA 02138, USA
\\
$^2$Department of Astronomy, Columbia University, 550 W 120th St, New York, NY 10027, USA}
\email{xinyi.guo@cfa.harvard.edu}
 \email{lsironi@astro.columbia.edu}
 \email{rnarayan@cfa.harvard.edu}
\begin{abstract} Recent X-ray observations of merger shocks in galaxy clusters have shown that the post-shock plasma is two-temperature, with the protons hotter than the electrons. By means of two-dimensional particle-in-cell simulations, we study the physics of electron irreversible heating in perpendicular low Mach number shocks, for a representative case with sonic Mach number of 3 and plasma beta of 16.
We find that two basic ingredients are needed for electron entropy production: (\textit{i}) an electron temperature anisotropy, induced by field amplification coupled to adiabatic invariance; and (\textit{ii}) a mechanism to break the electron adiabatic invariance itself. In shocks, field amplification occurs at two major sites: at the shock ramp, where density compression leads to an increase of the frozen-in field; and farther downstream, where the shock-driven proton temperature anisotropy generates strong proton cyclotron and mirror modes.  The electron temperature anisotropy induced by field amplification exceeds the threshold of the electron whistler instability. The growth of whistler waves breaks the electron adiabatic invariance, and allows for efficient entropy production. We find that the electron heating efficiency displays only a weak dependence on mass ratio (less than $\sim 30\%$ drop, as we increase the mass ratio from $m_i/m_e=49$ up to $m_i/m_e=1600$).  We develop an analytical model of electron irreversible heating and show that it is in excellent agreement with our simulation results. 
\end{abstract}


\maketitle

\section{Introduction} \label{sec:intro}
Galaxy clusters grow via mergers of subclusters. 
A large fraction of the kinetic energy of the infalling subclusters is dissipated at low Mach number  shocks ($M_s \lesssim 5$, where $M_s$ is the ratio of shock speed and pre-shock sound speed), that heat the intracluster medium (ICM) and sometimes accelerate particles to relativistic energies \citep{Sarazin2002,Ryu2003,Bruggen2012}.
Merger shocks in clusters are collisionless. Due to the high ICM temperatures ($\sim 10^7-10^8 \,$K) and low densities ($10^{-2} - 10^{-4}\, {\rm cm}^{-3}$), the collisional mean free path ($\sim 10^{21}-10^{23}\,$ cm) is as large as the size of the cluster.

Galaxy cluster shocks are routinely observed in the radio and X-ray bands.
X-ray measurements can quantify the density and temperature jumps between the unshocked (upstream) and the shocked (downstream) plasma \citep[e.g.,][]{Markevitch2002a,Finoguenov2010,Russell2010,Ogrean2013b,Eckert2016,Akamatsu2017}. The existence of shock-accelerated electrons is revealed by radio observations of synchrotron radiation \citep[e.g.,][]{VanWeeren2010,Lindner2014,Trasatti2015,Kale2017}.
Recently, the pressure jump associated with a merger shock at relatively high redshift has been measured through radio observations of the thermal Sunyaev-Zel'dovich (SZ) effect \citep{Basu2016}. 

Since all of our observational diagnostics are based on radiation emitted by electrons,  the proton properties (in particular, their temperature) are basically unconstrained. 
One usually makes the simplifying assumption that the electron temperature equals the mean gas temperature (and so, the proton temperature). This assumption is unlikely to hold in the vicinity of merger shocks. Ahead of the shock, the  bulk kinetic energy of protons is a factor of $m_i/m_e$ larger than for electrons (here, $m_i$ and $m_e$ are the proton and electron masses, respectively). In the absence of a channel for efficient proton-to-electron energy transfer, a comparable ratio should persist between the post-shock temperatures of the two species.

While Coulomb collisions will eventually drive electrons and protons to equal temperatures,  the collisional equilibration timescale  \citep{Spitzer1962} for typical ICM conditions is as long as  $10^{8}-10^9$ yrs. 
In fact, X-ray observations by \cite{Russell2012a} have shown that the electron temperature just behind a merger shock in Abell 2146 is lower than the mean gas temperature expected from the  Rankine-Hugoniot jump conditions, and thus lower than the proton temperature. As a separate evidence, 
\cite{Akamatsu2017} has compiled a list of merger shocks, estimating their Mach number from both X-ray ($M_{s,\rm X-ray}$) and radio observations ($M_{s,\rm radio}$), and noticed a slight bias of $M_{s,\rm radio}\gtrsim M_{s,\rm X-ray}$. Here, $M_{s,\rm radio}$ is derived by measuring the power-law slope of the synchrotron emission, which is related --- via the theory of  diffusive shock acceleration --- to the density compression at the shock (and so, to the Mach number). On the other hand, $M_{s,\rm X-ray}$ is obtained from the electron free-free emission by measuring the jumps in density and temperature across the shock. It follows that, if electrons have a lower temperature than protons behind the shock, $M_{s,\rm X-ray}$ would have been underestimated.

In fact, it has long been thought that collisionless shocks 
can lead to a two-temperature structure at the outskirts of galaxy clusters \citep{Fox1997,Ettori1998,Takizawa1999}. 
Detailed cosmological hydrodynamic simulations have shown that this can significantly bias the X-ray and thermal SZ signatures \citep{Wong2009,Rudd2009}. In the absence of a physical model for electron heating in low Mach number shocks, these studies usually employ an {\it ad-hoc} subgrid approach to prescribe the electron heating efficiency in shocks.  Either electrons are heated adiabatically, or the non-adiabatic (or ``irreversible'') heating efficiency is quantified by a phenomenological (and often, arbitrary) parameter. While observations from heliospheric low Mach number shocks have shown that electrons do not get heated much beyond adiabatic compression \citep{Bame1979,Ghavamian2013}, there has also been direct evidence of electron entropy production  (i.e., non-adiabatic heating) at  low Mach number shock fronts \citep{Parks2012}. 

What is then the mechanism responsible for electron heating at collisionless shocks? This is a fundamental  question of plasma physics, as the fluid-type Rankine-Hugoniot relations only predict the jump in the mean plasma temperature across the shock, without specifying how the shock-generated heat is distributed between the two species. 
To understand electron heating in collisionless shocks, fully-kinetic simulations with the particle-in-cell (PIC) method \citep{Birdsall1991,Hockney1981} are essential to self-consistently capture the non-linear structure of the shock and the role of electron and proton plasma instabilities in particle heating.

So far, PIC studies of electron heating in shocks have focused on the regime of high sonic Mach number ($M_s\gtrsim10$) and low plasma beta ($\beta_{p0}\lesssim 1$)  appropriate for supernova remnants. At very high Mach numbers, the Buneman instability can trap electrons in the shock transition region and heat them  \citep{Dieckmann2012}. 
For lower Mach numbers, resonant wave-particle scattering induced by the modified two-stream instability (MTSI) can lead  to significant electron heating at the shock front \citep{Matsukiyo2003,Matsukiyo2010}.

The regime of low sonic Mach number and high beta most relevant for cluster merger shocks is still unexplored.
In this paper, we study electron heating in low Mach number perpendicular shocks by means of two-dimensional (2D) PIC simulations. We focus on the results from a reference shock simulation with $M_s=3$ and $\beta_{p0}=16$. In a forthcoming paper (X. Guo et al., in preparation) we will explore the dependence of our conclusions on sonic Mach number and plasma beta. 
The choice of a perpendicular magnetic field geometry is meant to minimize the role of non-thermal electrons that are self-consistently accelerated in oblique configurations, as we have shown in  \citet{Guo2014,Guo2014c}. In the absence of shock-accelerated electrons returning upstream, the shock can settle to a steady state on a shorter time, thus allowing to focus on the steady-state electron heating physics. However,  we emphasize that we have verified with a suite of PIC simulations of quasi-perpendicular shocks  (not shown here) that the physics of electron heating presented in this paper also applies to quasi-perpendicular configurations, as long as the non-thermal electrons are energetically sub-dominant. 

The rest of the paper is organized as follows. 
In Section \ref{sec:phys} we lay out the theoretical framework for electron heating. 
Section \ref{sec:setup} describes the setup of the reference shock simulation. Section \ref{sec:shock} shows the shock structure of the reference simulation. We emphasize that efficient electron irreversible heating occurs at two main locations. With periodic box experiments meant to reproduce the shock conditions at the two major sites of entropy production,  Section \ref{sec:ramp} and Section \ref{sec:waves} investigate in detail the electron heating physics in these two locations,  and validate the heating theory presented in Section \ref{sec:phys}. In Section \ref{sec:back}, the heating model is then validated in the shock simulation. We conclude with a summary in Section \ref{sec:disc}. 

\section{$\!\!\!$ The Physics of Electron Heating}\label{sec:phys}
As they pass through the shock, electrons will experience a density compression, which results in adiabatic heating. In addition, irreversible processes might operate, which will further increase the electron temperature. The purpose of this section is to present a general formalism for the physics of irreversible heating. Even though we will be primarily interested in electron heating, the model can be applied to any particle species. It relies on the presence of two basic ingredients: (\textit{i}) a temperature anisotropy; and (\textit{ii}) a mechanism to break the adiabatic invariance. We first describe the change in internal energy of an anisotropic fluid, and then consider the resulting change in entropy.\footnote{We point out that the model that we propose is reminiscent of the so-called ``magnetic pumping'' mechanism, where a periodically-varying external magnetic field is used in the laboratory to drive proton anisotropy and subsequent plasma heating \citep{Spitzer1953,Berger1958,Borovsky1986}.}

\subsection{The Change in Internal Energy}
The work done on an isotropic gas with  pressure $P$ and volume $V$ is simply $dW=-PdV$. We shall generalize this expression to the case of an anisotropic gas having pressure perpendicular (parallel, respectively) to the magnetic field lines equal to $P_{\perp}$ ($P_\parallel$, respectively). Consider a magnetic flux tube with length
$L$, cross-sectional area $A$, volume $V=LA$, and field strength $B$. The magnetic flux through the area $A$ is $\Phi=BA$. In response to a compression (or expansion) perpendicular to the magnetic field, the volume will change as 
\be\label{eq:dVperp}
dV_{\perp} \!=\!L\,dA\!=\!L\,d\left(\frac{\Phi}{B}\right)\!=\!-L\Phi \frac{dB}{B^2}\!=\!-V  d\ln B~,
\ee
where we have used the fact that, because of flux freezing, $\Phi$ is a constant. In contrast, for compression (or expansion) along the field, the volume will change as 
\be\label{eq:dVpara}
dV_{\parr}\!=\!A\,dL\!=\!A\,d\!\left(\frac{V}{A}\right)\!=\!\frac{AN}{\Phi}\,d\!\left(\frac{B}{n}\right)\!=\!-Vd\ln\!\left(\frac{n}{B}\right)~,
\ee
where $N$ is the total number of particles in the volume element, with number density $n=N/V$. It follows that the work done on an anisotropic gas can be written as 
\be
d W & =&-P_{\perp} dV_\perp-P_\parr dV_\parr \nonumber\\
&=&P_{\perp}Vd\ln B+P_{\parallel}Vd\ln\left(\frac{n}{B}\right)~.
\ee
Defining the work done per particle as $dw=dW/N$, we find that it can be separated into a ``perpendicular'' component $d w_\perp$ and a ``parallel'' component $d w_\parr$ as
\begin{equation}\label{eq:worke}
d w=\underbrace{k_{\rm B}T_{\perp}d\ln B}_{d w_{\perp}}+\underbrace{k_{\rm B}T_{\parallel}d\ln\left(\frac{n}{B}\right)}_{d w_{\parallel}}~.
\end{equation}
It follows that $dw_\perp$ will change the internal energy per particle $u_\perp$ associated with motions perpendicular to the field, while $dw_\parallel$ will affect the energy per particle $u_\parr$ associated with parallel motions. 

In writing the energy equation for the perpendicular and parallel components, we need to take into account two additional processes: (\textit{i}) In the presence of pitch angle scattering, heat can be transferred between the two components (as we show below, this will give rise to entropy increase). We denote the differential amount of transferred heat as $dq_{\perp\to\parr}$, with the convention that $dq_{\perp\to\parr}>0$ if heat flows from the perpendicular to the parallel component. (\textit{ii}) Pitch angle scattering may be caused by self-generated waves (e.g., sourced by the plasma anisotropy), whose energy needs to be provided by the plasma itself. The energy balance relations then read
\begin{align}
d u_{\perp}&=dw_\perp-d q_{\perp\to\parallel}-d e_{w,\perp}\,\label{eq:duperp}\\
d u_{\parallel}&=dw_\parallel+d q_{\perp\to\parallel}-d e_{w,\parallel}\,\label{eq:dupara}
\end{align}
where we denote the wave energy per particle coming from the perpendicular (parallel, respectively) plasma energy as $de_{w,\perp}$ ($de_{w,\parallel}$, respectively). By summing the above two equations, we obtain the expected result that  the net change of internal energy per particle is equal to the external work minus the energy given to waves
\begin{equation}\label{eq:energyconserv}
d u \equiv d u_{\perp} + d u_{\parallel} = dw - d e_{w\,\rm tot} ~,
\end{equation}
where we denote the total energy per particle transferred to waves as $d e_{w\,\rm tot} \equiv  d e_{w,\perp} + d e_{w,\parallel}$, including magnetic, electric and bulk kinetic contributions (in practice, the magnetic term always dominates). 

While the total wave energy per particle $d e_{w\,\rm tot}$ is easy to extract from our simulations, the two  contributions $d e_{w,\perp}$ and $d e_{w,\parr}$ are hard to separate. We show below that for the entropy calculation it suffices to measure the total energy per particle transferred to waves $d e_{w\,\rm tot}$. We also remark that $d e_{w\,\rm tot}$ accounts for the differential energy per particle {\it transferred} to waves, which might not necessarily equal the differential change in the energy {\it residing} in waves, which we shall call $d e_{w}$. More specifically, while for electron-driven waves $d e_{w\,\rm tot}=d e_{w}$, we will show in \sect{waves} that proton-generated waves will lose energy by performing work on the electron plasma, so the change in the energy residing in proton waves $d e_{w}$ will be smaller than the differential energy $d e_{w\,\rm tot}$ transferred from protons to waves. 

\subsection{The Change in Entropy}
For a non-relativistic bi-Maxwellian plasma with perpendicular temperature $T_\perp$ and parallel temperature $T_\parallel$, the entropy per particle (or specific entropy) is 
\begin{equation}
s\equiv -\frac{\int d^3 p\, f\ln f}{\int d^3 p\, f} =\ln\left(\frac{T_{\perp}T_{\parallel}^{1/2}}{n}\right)+C~,\label{eq:bimaxs}
\end{equation}
where $f({\b p})$ is the phase space distribution and $C$ is a normalization constant. By differentiating, 
\begin{equation}
ds=\frac{dT_{\perp}}{T_{\perp}}+\frac{1}{2}\frac{dT_{\parallel}}{T_{\parallel}}-\frac{dn}{n}~.
\end{equation}
The temperature $T_{\perp,\parr}$ can be related to the internal energy per particle $u_{\perp,\parr}$ via the respective adiabatic index $\Gamma_{\perp,\parr}$ as 
\begin{equation}
u_{\perp,\parr}=\frac{k_{\rm B}T_{\perp,\parr}}{\Gamma_{\perp,\parr}-1}~.
\end{equation}
For a non-relativistic gas, $\Gamma_\perp=2$ (two degrees of freedom are available in the perpendicular direction), whereas $\Gamma_\parr=3$ (one degree of freedom). The  equation above then becomes
\begin{equation}
ds=\frac{du_{\perp}}{T_{\perp}}+\frac{du_{\parallel}}{T_{\parallel}}-\frac{dn}{n}~.
\end{equation}
Using Equations \eqref{eq:duperp} and \eqref{eq:dupara}, we have
\begin{align}
ds & = d q_{\perp\to\parallel}\left[\frac{1}{T_{\parallel}}-\frac{1}{T_{\perp}}\right]-\frac{d e_{w\perp}}{T_{\perp}}-\frac{d e_{w\parallel}}{T_{\parallel}}\,\label{eq:tosub}
\end{align}
which shows that the entropy of the gas can change as a result of heat flowing internally between the parallel and perpendicular components (first term on the right hand side) or when generating the waves (second term).
This can be rewritten in two equivalent forms:
\begin{align}
ds & = &\left[\frac{1}{2}d\ln\left(\frac{T_\parallel}{(n/B)^2}\right)\right]\cdot\left[1-\frac{T_\parallel}{T_{\perp}}\right]-\frac{d e_{w\,\rm tot}}{T_{\perp}}\,\label{eq:dselec}\\
ds &= & -\left[d\ln \left(\frac{T_{\perp}}{B}\right)\right]\cdot\left[\frac{T_{\perp}}{T_{\parallel}}-1\right]-\frac{d e_{w\,\rm tot}}{T_{\parallel}}.\label{eq:dsperp}
\end{align}
As anticipated above, the two separate components $de_{w,\parallel}$ and $de_{w,\perp}$ of the wave energy per particle do not explicitly enter the entropy equation. 

In Equations \eqref{eq:dselec} and \eqref{eq:dsperp}, the first term on the right hand side typically dominates. This clearly demonstrates that two ingredients are required for entropy generation: (\textit{i}) the presence of a temperature anisotropy; and (\textit{ii}) a mechanism to break the adiabatic invariance. Note that the CGL double adiabatic theory of \citet{Chew1956} predicts that, for adiabatic perturbations,  $T_\perp \propto B$ and $T_\parallel\propto (n/B)^2$, which follow from the conservation of the first and second adiabatic invariants. The form of Equations \eqref{eq:dselec} and \eqref{eq:dsperp} is thus easy to understand. In most cases, it is the temperature anisotropy that provides the free energy for generating the waves responsible for breaking the adiabatic invariance.

We conclude with an important remark on the nature of the magnetic field $B$. This should be meant as a large-scale field, so the particle response to its variation is properly modeled by the CGL approximation. In particular, the field that we have denoted as $B$ must not include the magnetic contribution of the waves that break the particle adiabatic invariance. In practice, $B$ will take into account all the magnetic contributions at scales much larger than the particle Larmor radius (for the species in question) and at frequencies much lower than the relevant gyration frequency. It follows that proton-generated waves that break the proton adiabatic invariance can still serve as large-scale $B$ fields for the electron energy and entropy equations, as we further discuss in \sect{waves}.

\section{Setup of the Shock Simulations}\label{sec:setup}
We perform numerical simulations using the three-dimensional (3D) electromagnetic PIC code TRISTAN-MP \citep{Spitkovsky2005}, which is a parallel version  of the code TRISTAN \citep{Buneman1993} that was optimized for studying collisionless shocks. 
In this section, we describe the setup of our shock simulations, which parallels closely what we used in \citet{Guo2014,Guo2014c}. In \sect{ramp} and in \sect{waves}, we will study in more detail the physics of electron heating by employing periodic simulation domains, meant to represent two specific regions of the shock structure. The simulation setups adopted there are different, and are described in the respective sections.

For shock simulations, we use a 2D simulation box in the $x-y$ plane, with periodic boundary conditions in the $y$ direction. Even though the simulations are two-dimensional in space, all three components of particle velocities and electromagnetic fields are tracked. The shock is set up by reflecting an upstream electron-proton plasma moving along the $-\hat{x}$ direction off a conducting wall at the leftmost boundary of the computational box ($x=0$). The interplay between the reflected stream and the incoming plasma causes a shock to form, which propagates along $\hat{x}$. In the simulation
frame, the downstream plasma is at rest.

The upstream electron-proton plasma is initialized following the
procedure described by \citet{Zenitani2015a}, as a drifting
Maxwell-J\"uttner distribution with electron temperature $T_{e0}$
equal to the proton temperature $T_{i0}$ (i.e. $T_{e0}=T_{i0}=T_0$),
and bulk velocity ${\b V}_0 =-V_0\hat{x}$. This gives a
simulation-frame Mach number \be
M_{s,0}=\frac{V_0}{c_s}=\frac{V_0}{\sqrt{2 \Gamma k_{\rm B}
    T_0/m_i}}~~,\label{eq:mach0} \ee where $c_s$ is the sound speed in
the upstream, $k_{\rm B}$ is the Boltzmann constant, $\Gamma=5/3$ is
the adiabatic index for an isotropic non-relativistic gas, and $m_i$
is the proton mass. Below, we will adopt the usual definition of Mach
number, as the ratio between the upstream flow velocity and the
upstream sound speed in the shock rest frame (rather than in the
downstream frame of the simulations, as in \eq{mach0}), where the
upstream moves into the shock with speed $V_1$. We will then
parameterize our results with the Mach number \be
M_s=\frac{V_1}{c_s}~~.  \ee The shock-frame Mach number $M_s$ is
related to the downstream-frame Mach number $M_{s,0}$ via \be
M_{s}=M_{s,0} \left(1+\frac{1}{r(M_s)-1}\right)~~, \ee where the
density jump $r(M_s)$ across the shock, in the limit of weakly
magnetized flows, is equal to \be
r(M_s)=\frac{\Gamma+1}{\Gamma-1+2/M_s^2}~~.  \ee In writing these
relations we have assumed an isotropic gas, which is valid upstream of
the shock by our initial conditions, and is also valid sufficiently
downstream of the shock, as we will see in the discussion that
follows. 

The incoming plasma carries a uniform magnetic field ${\b B}_0$, and its associated motional electric field ${\b E}_0=-{\b V}_0/c\cross {\b B}_0$. The magnetic field strength  is parametrized by the plasma beta 
\begin{equation}
\beta_{p0}=  \frac{8\pi n_{0} k_{\rm B} (T_{i0}+T_{e0})}{B_0^2}=\frac{16\pi n_{0} k_{\rm B} T_0}{B_0^2}~~,
\end{equation}
where $n_{i0}= n_{e0}= n_0$ is the number density of the incoming protons and electrons. Alternatively, one could quantify the magnetic field strength via the Alfv\'enic Mach number $M_A=M_s \sqrt{\Gamma \beta_{p0}/2}$. 

The magnetic field is initialized in the simulation plane along the  $y$ direction, i.e.,  perpendicular to the shock normal. 
We find that the shock physics is properly captured by 2D simulations only if the field is lying in the
simulation plane. A posteriori, this will be motivated by the fact that the plasma instabilities excited in the downstream region have wavevectors preferentially parallel or quasi-parallel to the background magnetic field. We have explicitly verified with an additional simulation having magnetic field initialized along $z$ (so, pependicular to both the shock normal and the simulation plane), that the electron heating efficiency is completely suppressed, just as in 1D simulation results (\app{outplane}). Our choice of an in-plane magnetic field configuration will be justified again in the following sections.

For accuracy and stability, PIC codes have to resolve the plasma oscillation frequency of the electrons 
\begin{equation}
\omega_{pe} = \sqrt{4\pi e^2n_0/m_e}~,
\end{equation} 
and the electron plasma skin depth $c/\omega_{pe}$, where $e$ and $m_e$ are the electron charge and mass, respectively. On the other hand, the shock structure is controlled by the proton Larmor radius 
\begin{equation}\label{eq:rLi}
r_{{\rm Li}}=M_{A,0}\sqrt{\frac{m_i}{m_e}}\ \frac{c}{\omega_{pe}}\gg \frac{c}{\omega_{pe}}~,
\end{equation} 
where the shock-frame \alf ic Mach number is $M_{A,0}=M_{s,0} \sqrt{\Gamma \beta_{p0}/2}$. Similarly, the evolution of the 
shock occurs on a time scale given by the proton Larmor gyration period $\Omega_{ci}^{-1}=r_{\rm Li}V_0^{-1}\gg \omega_{pe}^{-1}$.
The need to resolve the electron scales, and at the same time to capture the shock evolution for many $\Omega_{ci}^{-1}$, is an enormous computational challenge for the realistic mass ratio $m_i/m_e=1836$. Thus we adopt a reduced mass ratio $m_i/m_e=49$ for our reference run, which is sufficient to properly separate the electron and proton scales. This allows us to follow the system for long times, until the shock reaches a steady state. We have explicitly verified that the electron heating physics in our shock simulations is nearly the same for higher mass ratios (see \sect{back}, where we test up to $m_i/m_e=200$). In addition, in \sect{ramp} and \sect{waves} we demonstrate via analytical arguments and PIC simulations in periodic domains that the electron heating efficiency is nearly independent of $m_i/m_e$ over the range from $m_i/m_e=49$ up to the realistic mass ratio.

As in \citet{Guo2014,Guo2014c},
the upstream plasma is initialized at a ``moving injector", which recedes from the wall in the $+\hat{x}$ direction at the speed of light. When the injector approaches the right boundary of the computational domain, we expand the box in the $+\hat{x}$ direction. 
This way both memory and computing time are saved, while following at all times the evolution of the upstream regions that are causally connected with the shock. Further numerical optimization can be achieved by allowing the moving injector to periodically jump backward (i.e. in the $-\hat{x}$ direction), resetting the fields to its right (see \cite{Sironi2009a}). For a perpendicular shock (i.e., with magnetic field perpendicular to the shock direction of propagation), no particles are expected to escape ahead of the shock, so we choose to jump the injector in the $-\hat{x}$ direction so as to keep a distance of a few proton Larmor radii ahead of the shock.  This suffices to properly capture the heating physics of electrons and protons.
We have checked, though only for relatively early times, that simulations with and without the jumping injector give identical results. 

In the main body of this paper, we present the results from a reference run with $M_s = 3$ and $\beta_{p0}=16$, as motivated 
by galaxy cluster shocks. The upstream plasma is initialized with $T_{i0}=T_{e0} = 10^{-2} m_e c^2$
and $V_0 = 0.05c$. We remark that even though our values for the plasma temperature and bulk speed are motivated by galaxy cluster shocks, the results can be readily applied to other systems (e.g., the solar wind), as long as the dimensionless ratios $M_s$ and $\beta_{p0}$ are the same and all the velocities remain non-relativistic. We will investigate the dependence of the results on the Mach number and the plasma beta in a forthcoming paper (X. Guo et al., in preparation).

We employ a spatial resolution of $10$ computational cells per electron skin depth $c/\omega_{pe}$, which is sufficient to resolve the Debye length of the upstream electrons for our chosen temperature of $k_{\rm B}T_{e0}= 10^{-2}m_e c^2 $. We have tested that a spatial resolution of 7 cells per electron skin depth can still capture  the electron heating physics. We use a time resolution of $dt = 0.045\ \omega_{pe}^{-1}$. 
Each cell is initialized with $32$ computational particles ($16$ per species), but we have tested that a larger number of particles per cell (up to 64 per species) does not change our results (\app{ppc}).  
For the reference run, the transverse size of the computational box is $151\ c/\omega_{pe}$, corresponding to $\sim 3\rli$, but  we have tested that simulations with a transverse box size up to $256\ c/\omega_{pe}\sim 5\rli$ show essentially the same results.

\begin{figure}[tbh]
\begin{center}
\includegraphics[width=0.5\textwidth]{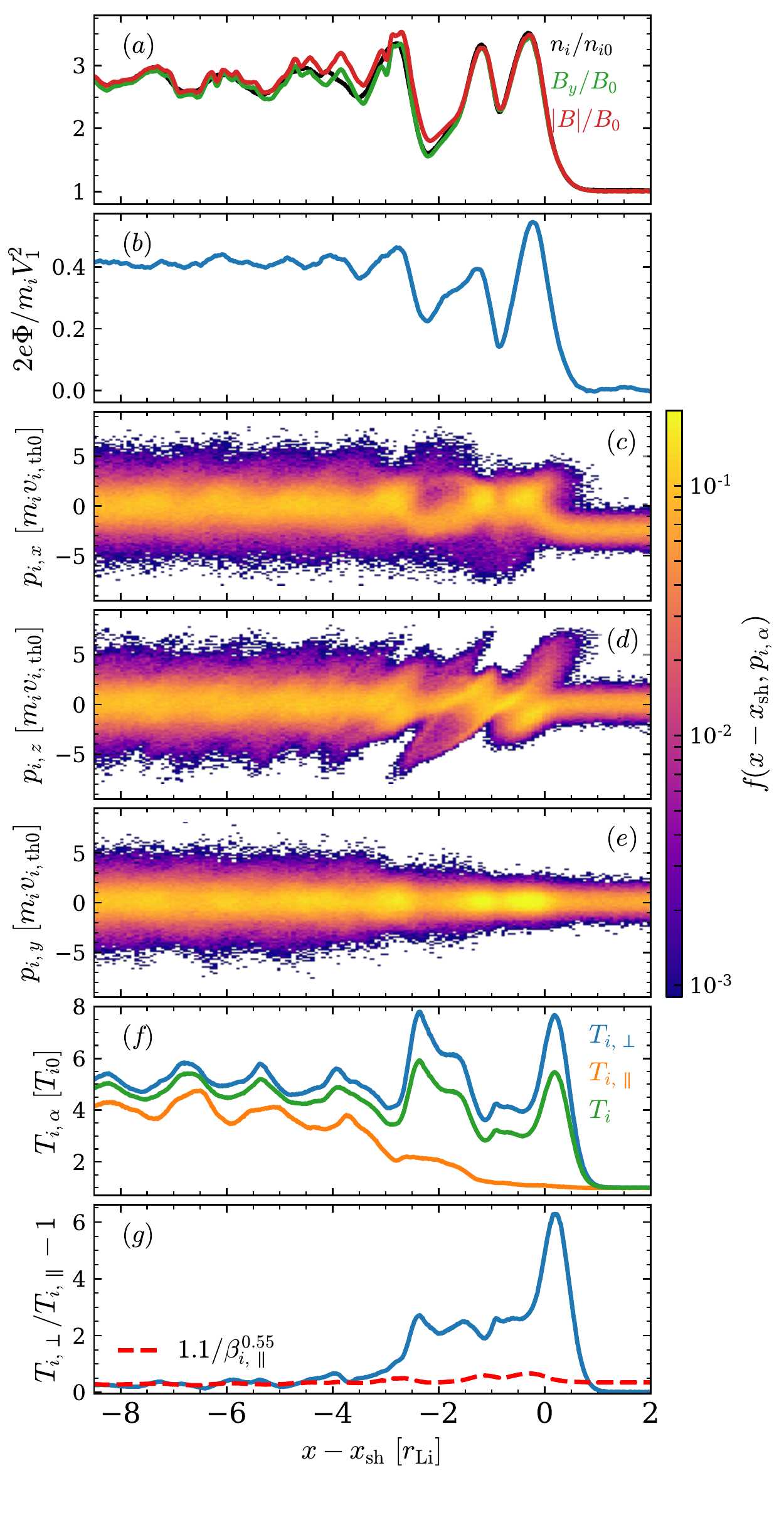}
\end{center}
\caption{Shock structure and proton dynamics at $t  = 25.6\, \Omega_{ci}^{-1}$. The $x$ coordinate is measured relative to the shock location $x_{\rm sh}$, and it is normalized to the proton Larmor radius $\rli$. From top to bottom, we plot: (a) the $y$-averaged 1D profiles of proton density (black, in units of the upstream value),  magnetic field $B_y$ (green, in units of the upstream field $B_0$) and total magnetic field strength $B$ (red, in units of the upstream field $B_0$); (b) the cross-shock electrostatic potential energy $e\Phi$, in units of the proton upstream bulk energy $m_i V_1^2/2$; (c)-(e) the proton phase spaces $f(\xsh,p_{i,x})$, $f(\xsh,p_{i,z})$, and $f(\xsh,p_{i,y})$, where the proton momentum $p_{i,\alpha}$ is in units of $m_i v_{i,\rm th 0}$ and the proton thermal velocity is defined as $v_{i,\rm th0}=\sqrt{k_{\rm B} T_{i0}/m_i}$; (f) the proton temperature perpendicular ($T_\perpi$, blue line) and parallel ($T_\pari$, orange line) to the magnetic field, and the mean proton temperature $T_i\equiv(2T_\perpi+T_\pari)/3$ (green line); (g) the proton anisotropy $T_\perpi/T_\pari-1$ (blue line) and the anisotropy upper bound in \eq{ionthresh} (red dashed line).
}\label{fig:protons}
\end{figure}

\begin{figure}[tbh]
\begin{center}
\hspace{0cm}
\includegraphics[width=0.55\textwidth]{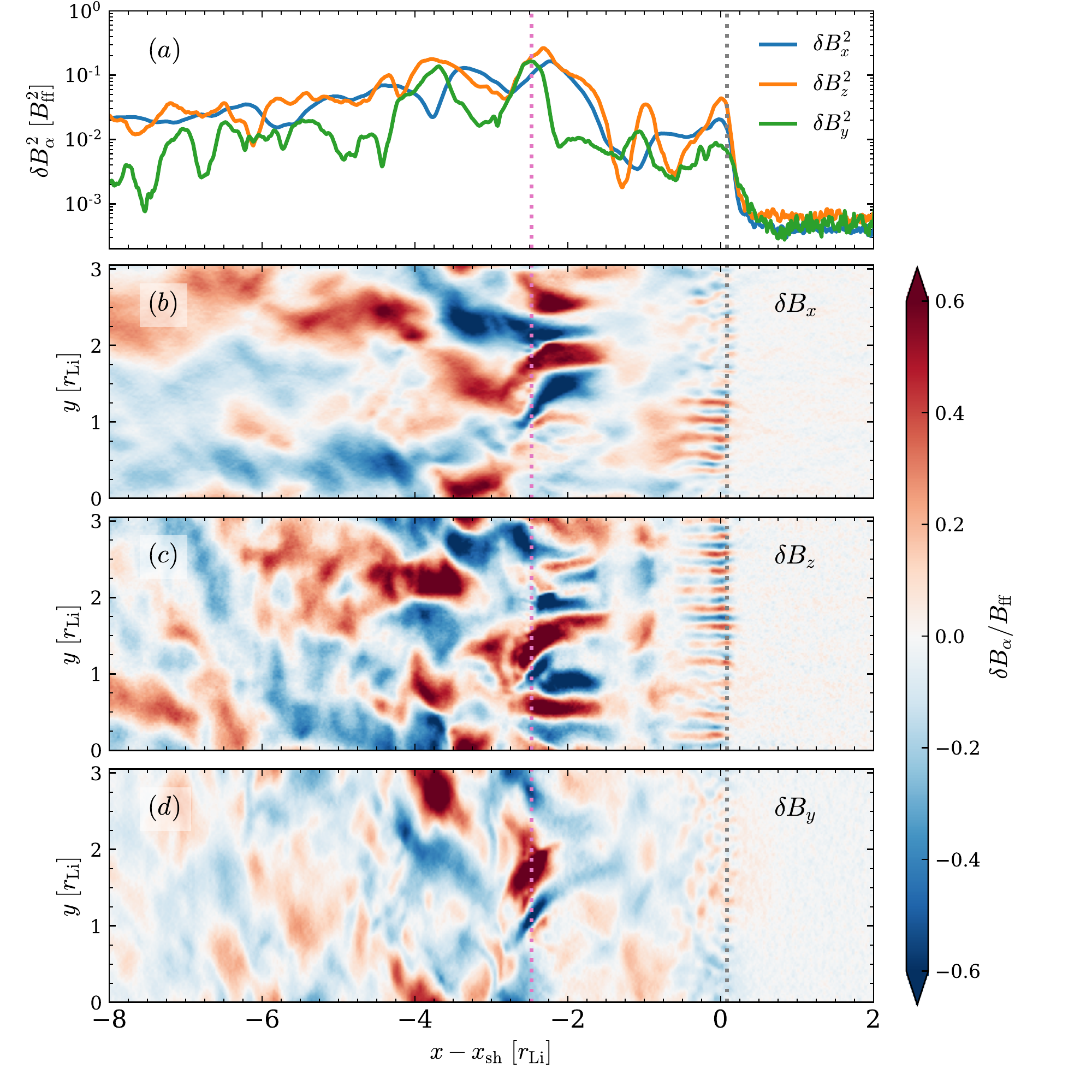}
\end{center}
\caption{1D and 2D structure of magnetic fluctuations in our reference shock run at $t  = 25.6\, \Omega_{ci}^{-1}$. In panel (a), we plot the energy of magnetic field fluctuations in the $x$, $z$ and $y$ directions (blue, orange and green lines, respectively) normalized to the magnetic energy of the frozen-in field, which is defined as ${\b B}_{\rm ff}\equiv B_{\rm ff}\hat{y}\equiv B_0 (n/n_0)\hat{y}$. Panels (b)-(d) show the 2D structure of the field fluctuations $\dbx=B_x/B_{\rm ff}$, $\dbz=B_z/B_{\rm ff}$ and $\dby=(B_y-B_{\rm ff})/B_{\rm ff}$, respectively. The $x$ coordinate is measured relative to the shock location $x_{\rm sh}$, and it is normalized to the proton Larmor radius $\rli$. In panels (b)-(d), the $y$ coordinate is in units of the proton Larmor radius $\rli$.}\label{fig:fields}
\end{figure}

\section{Shock Structure}\label{sec:shock}
In this section, we describe the structure of our reference shock run, with $M_s=3$, $\beta_{p0}=16$ and $m_i/m_e=49$. We first discuss the proton dynamics and the generation of magnetic fluctuations sourced by the proton temperature anisotropy. Then, we present the electron dynamics and focus on the profile of electron irreversible heating. We will identify two main locations where the electron entropy increases: the shock ramp and the site where proton-driven waves grow in the downstream. The electron heating physics in these two regions will be investigated in \sect{ramp} and \sect{waves}, respectively.

\subsection{Proton Dynamics and Proton-Driven Instabilities}\label{sec:protons}
In this subsection, we describe the proton dynamics, with a focus on proton isotropization and thermalization downstream of the shock.
Figure \ref{fig:protons} shows the profile of various quantities in the shock at time $t=25.6\, \Omega_{ci}^{-1\
}$, as a function of the $x$ coordinate relative to the shock location $x_{\rm sh}$, in units of the proton Larmor radius $\rli$ defined in \eq{rLi}.

Panel (a) shows the $y$-averaged profile of the proton number density $n_i$ in units of the proton density in the upstream $n_{i0}$ (black line). The density compression at the shock reaches $n_i/n_{i0}\sim 3.5$ over a distance of $\sim r_{\rm Li}$, consistent with the expectation that the thickness of a perpendicular shock should be of the order of the proton Larmor radius \citep{Bale2003,Scholer2006a}.
The density oscillates on a typical length scale of $\sim r_{\rm Li}$ after the overshoot
and then relaxes to the Rankine-Hugoniot value of $\sim 2.8$ beyond a distance of $\sim 5\, r_{\rm Li}$ behind the shock. 

The density pile-up at the shock is related to the electrostatic potential $\Phi$ that develops in the shock transition region. This phenomenon has been well studied via hybrid simulations of collisionless shocks \citep[e.g.][]{Leroy1981,Leroy1982,Leroy1983}.  As shown in \fig{protons}(b), the potential energy $e\Phi$ reaches $\sim 60\%$ of the incoming proton energy $m_i V_1^2/2$. As a result, a significant fraction of the incoming protons are reflected back toward the upstream, leading to a pile-up of particles just in front of the shock. The reflected protons can be identified as the ones with positive $p_{i,x}$ and $p_{i,z}$ ahead of the shock in the phase spaces of \fig{protons}(c) and (d), respectively. As the reflected protons gyrate in the shock-compressed magnetic field, they gain energy from the upstream motional electric field. 
Upon their second encounter with the cross-shock potential, the reflected protons now have sufficient energy to penetrate the shock. In the downstream region just behind the shock, the protons keep gyrating in the $xz$ plane perpendicular to the shock-compressed magnetic field (compare the phase spaces in \fig{protons}(c) and (d), at $-4\lesssim (\xsh)/\rli\lesssim 0$). The peaks in density seen in \fig{protons}(a) are then correlated with the locations where the proton gyro-phase is such that most protons have small $p_{i,x}$ (e.g., at $\xsh\sim -0.25 \rli$, $-1.25\rli$ and $-2.75\rli$). The amplitude of the density oscillations gets smaller as the gyrating reflected  protons become more and more phase-mixed with the directly transmitted protons, at $\xsh\lesssim -5 \,r_{\rm Li}$.

Since the post-shock protons gyrate in the $xz$ plane perpendicular to the field, the momentum dispersion along the $y$ direction of the field is expected to be nearly the same on the two sides of  the shock (see the $p_{i,y}$ phase space in \fig{protons}(e) near the shock). Further behind the shock, the dispersion in $p_{i,y}$ increases. This can be also quantified with the $y$-averaged profiles of the proton temperature perpendicular ($T_{i,\perp}$) and parallel ($T_{i,\parallel}$) to the background magnetic field, as in \fig{protons}(f). Here, the $jk$ component of the temperature tensor is defined as $k_{\rm B}T_{jk}/m_i c^2 \equiv \langle \gamma' v_j' v_k'\rangle/ c^2$, where $v_j',v_k'$ are the particle velocities in the fluid comoving frame, $\gamma'$ is the comoving particle Lorentz factor, and the average is performed over the particle distribution at a given spatial location. As \fig{protons}(f) shows, the mean proton temperature $T_i$, defined as\footnote{The factor of two that multiplies $T_\perpi$ in the definition of $T_i$ comes from the fact that the perpendicular motion has two degrees of freedom.}
\be
T_i=\frac{2 \,T_{i,\perp}+T_{i,\parr}}{3}~,
\ee
is nearly uniform in the downstream region (green line), but the parallel temperature (orange line) --- which is continuous across the shock --- increases with distance behind the shock, while the perpendicular temperature (blue line) shoots up at the shock and then experiences a modest decline. This is the same trend shown by the phase spaces in \fig{protons}(c)-(e).

The decrease in perpendicular temperature, and the resulting increase in parallel temperature, suggests that protons are being scattered in pitch angle. In fact, in the region $-4 \lesssim (\xsh)/\rli\lesssim -1$ where the variation in $T_\perpi$ and $T_\pari$ is most pronounced, strong magnetic waves are observed in \fig{fields}. Their wavelength is comparable to the proton skin depth, indicating that they are driven by protons (as opposed to electrons). In \fig{fields}(a), we compare the 1D profiles (averaged over the $y$ direction) of the magnetic fluctuations $\dbx^2$, $\dby^2$ and $\dbz^2$, normalized to $B_{\rm ff}^2$, where $B_{\rm ff}$ is defined as the magnitude of the flux-frozen magnetic field (i.e., ${\b B}_{\rm ff}\equiv B_0 (n/n_0)\hat{y}$, where $n$ is the $y$-averaged particle density).\footnote{The frozen-in magnetic field is also used in the definition of $\dby\equiv B_y-B_{\rm ff}$.} The energy of proton-driven waves peaks at $\xsh\sim-2.5\rli$. In \fig{protons}(a), they are responsible for the excess of magnetic field strength (red curve) above the flux-freezing prediction (which would correspond to the density profile, in black).

The dominant mode at $-4\lesssim\xshnorm\lesssim-1$ in the $x$ and $z$ direction  has a wavevector nearly parallel to the background field (\fig{fields}(b) and (c)), consistent with the proton cyclotron instability \citep{Kennel1966,Davidson1975}. The waves in $\dby$ are slightly weaker (compare the green line with the blue and orange curves in \fig{fields}(a)) and have oblique wavevectors (\fig{fields}(d)), as expected for the mirror mode \citep{Chandrasekhar1958,Barnes1966,Hasegawa1975,Mckean1993}. The presence of mirror modes breaks the flux freezing condition, as shown by the fact that in \fig{protons}(a) the $y$-averaged transverse magnetic field profile $B_y/B_0$ deviates at $-5\lesssim \xshnorm \lesssim -2$ from the density profile (in black, which tracks the flux freezing prediction).

Both the proton cyclotron instability and the mirror instability are sourced by proton temperature anisotropy. In fact, since the motion of downstream protons right behind the shock is mostly confined in the $xz$ plane, a large temperature anisotropy arises, with $T_\perpi\gg T_\pari$ (\fig{protons}(g)). The anisotropy provides free energy for the growth of proton cyclotron waves and mirror modes, which scatter the protons in pitch angle and reduce their anisotropy back to the upper bound corresponding to marginal stability \cite{Gary1997} (see the red dashed line in \fig{protons}(g)), which is
\be\label{eq:ionthresh}
\frac{T_\perpi}{T_\pari}-1\simeq \frac{1.1}{\beta_\pari^{0.55}}~.
\ee
Here, $\beta_\pari=8 \pi n_i k_{\rm B} T_\pari /B^2$ is the local value of the proton plasma beta, computed with the parallel proton temperature.

\begin{figure}[tbh]
\begin{center}
\includegraphics[width=0.475\textwidth]{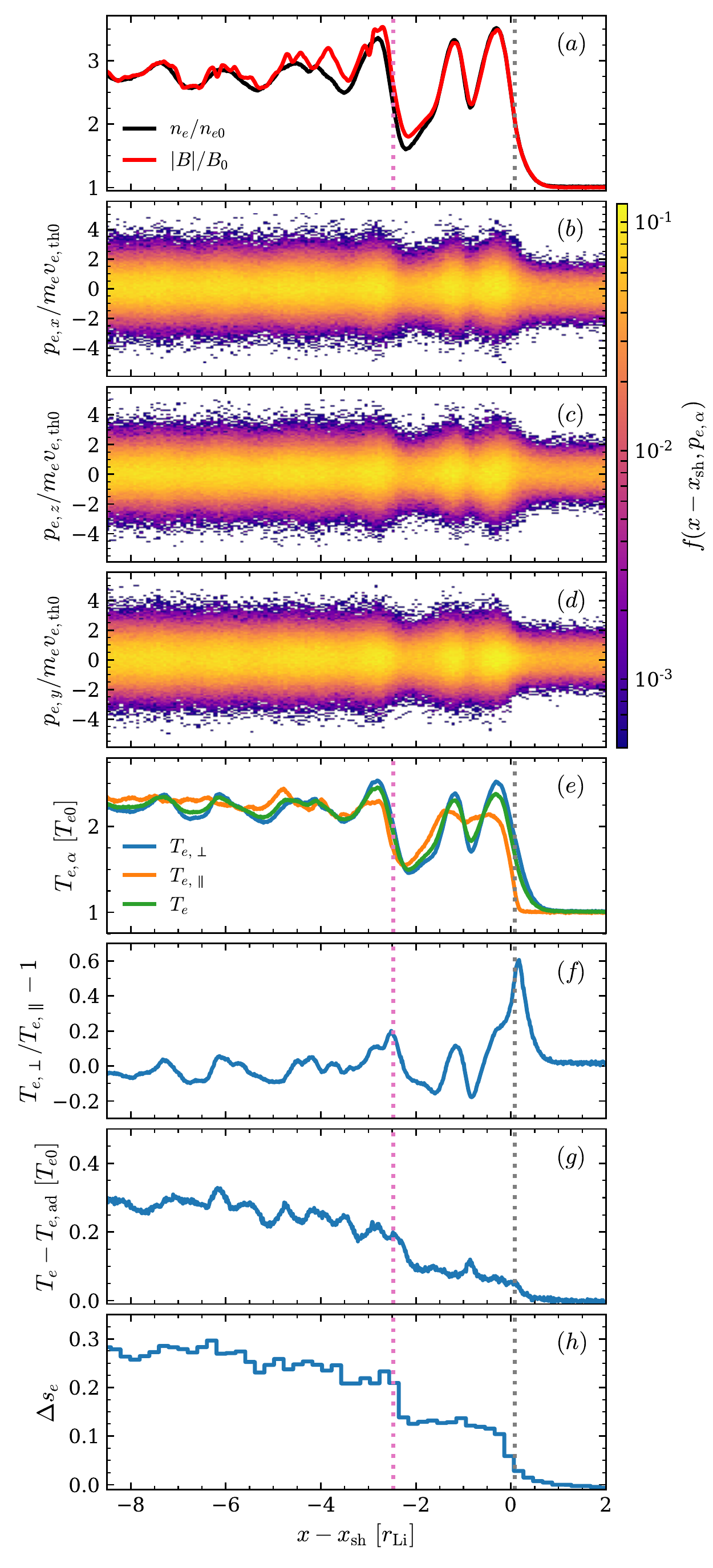}
\end{center}
\caption{Shock structure and electron dynamics at $t  = 25.6\, \Omega_{ci}^{-1}$. From top to bottom, we plot: (a) the $y$-averaged profiles of electron density (black) and total magnetic field strength $B$ (red); (b)-(d) the electron phase spaces $f(\xsh,p_{e,x})$, $f(\xsh,p_{e,z})$, and $f(\xsh,p_{e,y})$, where the electron momentum $p_{e,\alpha}$ is in units of $m_e v_{e,\rm th 0}$ and the electron thermal velocity is  $v_{e,\rm th0}=\sqrt{k_{\rm B} T_{i0}/m_i}$; (e) the electron temperature perpendicular ($T_\perpe$, blue) and parallel ($T_\pare$, orange) to the magnetic field, and the mean electron temperature $T_e\equiv(2T_\perpe+T_\pare)/3$ (green); (f) the electron anisotropy $T_\perpe/T_\pare-1$; (g) the excess electron temperature $T_e$ over the adiabatic expectation $T_{e,\rm ad}=(n_e/n_{e0})^{2/3}T_{e0}$ for an isotropic gas; (h) the electron entropy profile, measured as in \eq{sdef}.}\label{fig:elecs}
\end{figure}

\subsection{Electron Dynamics and Heating}\label{sec:lecs}
In this subsection, we describe the electron dynamics, with a focus on electron heating in the shock layer and in the downstream region. 
Due to their opposite charge and much smaller Larmor radius, 
the  dynamics of electrons is drastically different from that of the protons. 

Figure \ref{fig:elecs}(a) shows the electron density profile (black line), which strongly resembles that of the protons (black line in Figure \ref{fig:protons}(a)), and thus ensures approximate charge neutrality. While a small degree of charge separation at the shock is responsible for establishing the electric potential $\Phi$ shown in \fig{protons}(b), the fact that $\Phi$ is nearly uniform at $\xsh\lesssim -5 \rli$ suggests that charge neutrality is satisfied very well in the far downstream.

Figure \ref{fig:elecs}(b)-(d) shows the electron phase space. Since electrons have opposite charge than protons, they are not reflected back upstream by the cross-shock potential. In fact, unlike for protons, there is no reflected electron population with $p_{e,x}\gtrsim 0$ just ahead of the shock (compare \fig{elecs}(b) with \fig{protons}(c)). 

Figure \ref{fig:elecs}(e) shows the temperature profile of electrons, for the perpendicular component $T_\perpe$ (blue), the parallel component $T_\pare$ (orange) and the mean temperature $T_e$ (green), which is defined as
\be
T_e=\frac{2 \,T_{e,\perp}+T_{e,\parr}}{3}~.
\ee
The profile of perpendicular temperature (blue line) follows closely the density compression (compare with the black line in \fig{elecs}(a)) and starts to rise just ahead of the shock at $\xsh\sim 0.5 \,r_{\rm Li}$. This is consistent with the double adiabatic theory \cite{Chew1956}, which predicts $T_{\perp}\propto B$ (and in flux freezing, $B\propto n$).\footnote{We remark, as we have already pointed out at the end of \sect{phys}, that the field strength $B$ should include all the magnetic contributions at scales much larger than the electron Larmor radius and at frequencies much lower than the electron gyration frequency.} The double adiabatic theory applies to electrons, since the density and magnetic field compression occurs on scales much larger than the electron Larmor radius. This is not  true for protons, since the shock thickness and the scale length of the downstream oscillations seen in \fig{protons}(a) are set by the proton Larmor radius $\rli$. 

The parallel electron temperature (orange line in \fig{elecs}(e)) initially remains unchanged, as the CGL theory predicts $T_{\parallel}\propto (n/B)^2$ and $B\propto n$ as a result of flux freezing  (compare the green and black lines in Figure \ref{fig:protons}(a) in the vicinity of the shock).
The increase in perpendicular temperature at the shock, while the parallel temperature stays the same as in the upstream, leads to a strong electron anisotropy, up to $T_{e,\perp}/T_{e,\parallel}-1\sim 0.6$ (\fig{elecs}(f)).
This excites the electron whistler instability, which creates the small-wavelength transverse magnetic waves in $\dbx$ and $\dbz$ seen in the region $\xsh\sim  -0.25\, r_{\rm Li} $ of Figure \ref{fig:fields}(b) and (c) (see also the magnetic energy in $\dbx^2$ and $\dbz^2$ in \fig{fields}(a), at the same location).  The electron whistler instability provides a mechanism for electron pitch angle scattering and thus reduces the electron temperature anisotropy, as shown in the downstream region of \fig{elecs}(f). 

As we have already discussed, if the electron fluid were to follow the double adiabatic predictions, $T_\perpe\propto n$ and $T_\pare\propto$ const. The fact that the perpendicular temperature profile in \fig{elecs} (f) (blue line) resembles the density profile (black line in \fig{elecs}(a)), and the fact that $T_e\sim T_\perpe$ (compare green and blue curves in \fig{elecs}(e)), suggests that most of the increase in electron temperature comes from adiabatic compression. However, the fact that $T_\pare$ is not constant across the shock requires non-adiabatic processes. In order to quantify the degree of non-adiabatic (or, ``irreversible'') electron heating, we compare in \fig{elecs}(g) the mean electron temperature $T_e$ with the adiabatic prediction
\begin{equation}
\frac{T_{e,\rm ad}}{T_{e0}} = \left(\frac{n_e}{n_{e0}}\right)^{2/3}.
\end{equation}
This estimate of the adiabatic temperature assumes an isotropic gas, which is valid, given the small degree of electron anisotropy far downstream of the shock (see \fig{elecs}(f) at $\xsh\lesssim -1\rli$).
Figure \ref{fig:elecs}(g) shows the excess of $T_e$ above $T_{e,\rm ad}$ in units of the upstream electron temperature. Most of the irreversible heating occurs at two locations: $\xsh\sim0$, i.e., in the shock transition region; and $\xsh\sim -2.5\, r_{\rm Li}$, where the density suffers another compression, and strong proton-driven waves are generated (see \fig{fields}). These two locations are marked by the vertical dotted lines in \fig{fields} and \fig{elecs}, and the particle and wave properties there will be further studied below.
In the far downstream, the  temperature excess over the adiabatic estimate saturates at $T_e-T_{e,\rm ad}\sim 0.3\, T_{e0}$ (\fig{elecs}(g)). 

An alternative (and possibly, more rigorous) estimate of the degree of irreversible electron heating is given by the specific entropy $s_e$ (i.e., the entropy per particle), measured with the electron distribution function $f_e$ as 
\begin{equation}\label{eq:sdef}
s_e \equiv -\frac{\int d^3p\  f_e\ln  f_e}{\int d^3p\ f_e}\,
\end{equation}
where the normalization is such that $\int d^3p\, f_e = n_e/n_{e0}$. 
To construct the spatial profile of $s_e(x)$, we first bin the particles by their $x$ position with a width of $\Delta x = 100$ cells. In each spatial bin, we compute $f_e({\b p})$ by constructing a three-dimensional histogram of the particle momenta. In each direction (i.e., $p_{e,x}$, $p_{e,y}$ and $p_{e,z}$), the central bin of the histogram lies at the mean momentum, and the histogram spans four standard deviations above and below the mean. Each standard deviation is resolved with 10 momentum bins. 

Figure \ref{fig:elecs}(h) shows the change of electron entropy with respect to the upstream value. 
In analogy to Figure \ref{fig:elecs}(g), the increase in electron entropy is localized around $\xsh\sim 0$ and $\xsh\sim-2.5\ r_{\rm Li}$ (indicated by the gray and pink vertical dotted lines, respectively). The increase in  electron specific entropy saturates at $\Delta s_e\sim 0.25$ in the far downstream. 

\begin{figure*}[tbh]
\begin{center}
\includegraphics[width=0.9\textwidth]{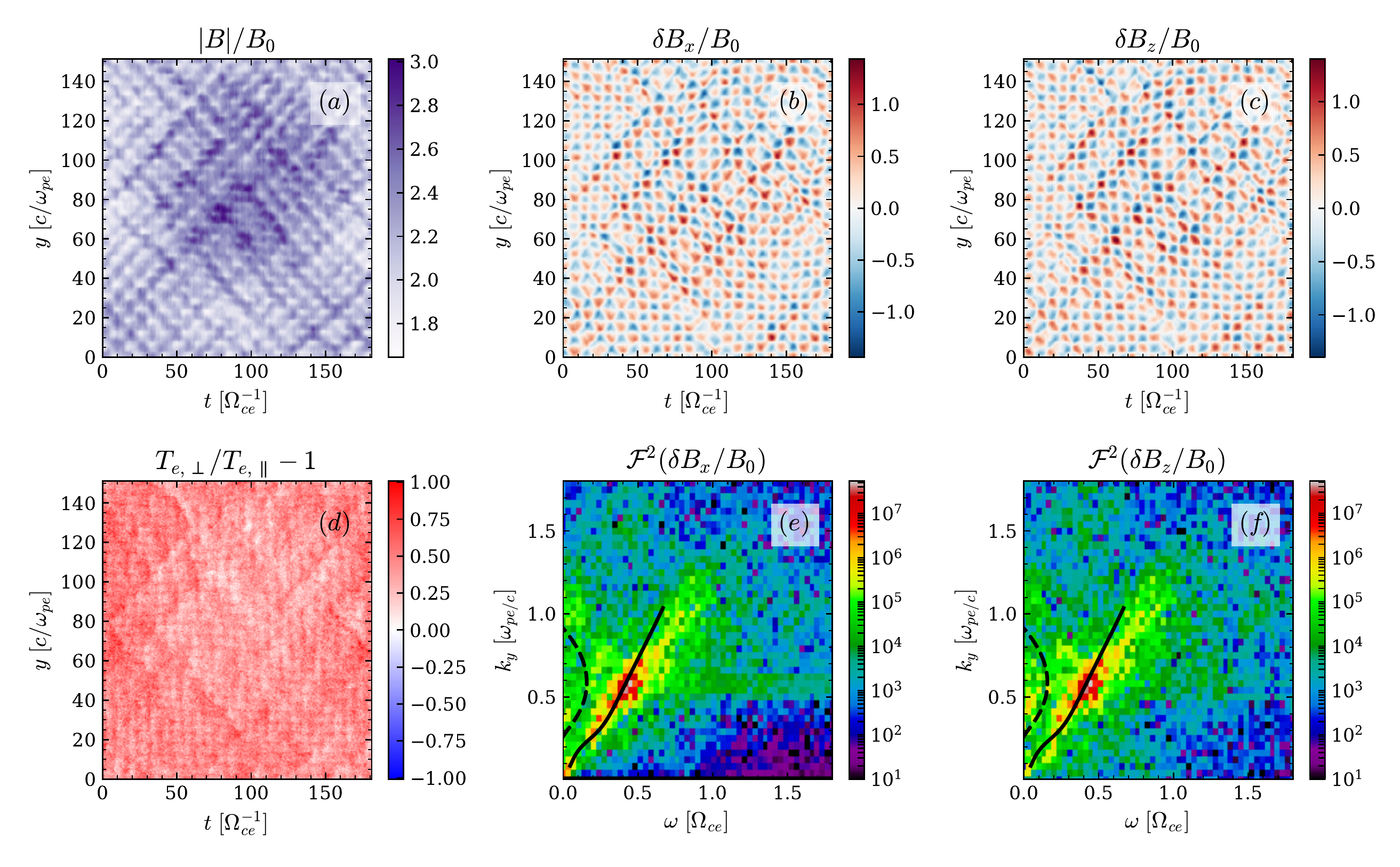}
\end{center}
\caption{Space-time diagrams and power spectra at a distance of $\xsh=4\, c/\omega_{pe}$ ahead of the shock (as indicated by the vertical dotted grey lines in \fig{fields} and \fig{elecs}), during the time interval $24.0\leq\Omega_{ci}t\leq27.4$. For this plot, the unit of time is the electron cyclotron time $\Omega_{ce}^{-1}$ (the corresponding unit of frequency is $\Omega_{ce}$) and the unit of distance along $y$ is the electron skin depth $\compe$ (the corresponding unit for the wavevector $k_y$ is $\omega_{pe}/c$). Panels (a)-(d) are the space-time diagrams of: (a) total magnetic field strength $|B|$, in units of the upstream field $B_0$; (b)-(c) transverse magnetic field fluctuations $\dbx/B_0$ and $\dbz/B_0$; (d) electron anisotropy $T_\perpe/T_\pare-1$. Panels (e) and (f) show the $(\omega,k_y)$  power spectra of the field fluctuations presented in panels (b) and (c), respectively. In panels (e) and (f), the solid black line is the predicted real part of the frequency of electron whistler modes, whereas the dashed black line is the predicted imaginary part (i.e., the growth rate). The agreement between the prediction and our measurement confirms that the fluctuations in panels (a)-(c) are whistler waves.}\label{fig:cut1}
\end{figure*}

\begin{figure*}[tbh]
\begin{center}
\includegraphics[width=0.9\textwidth]{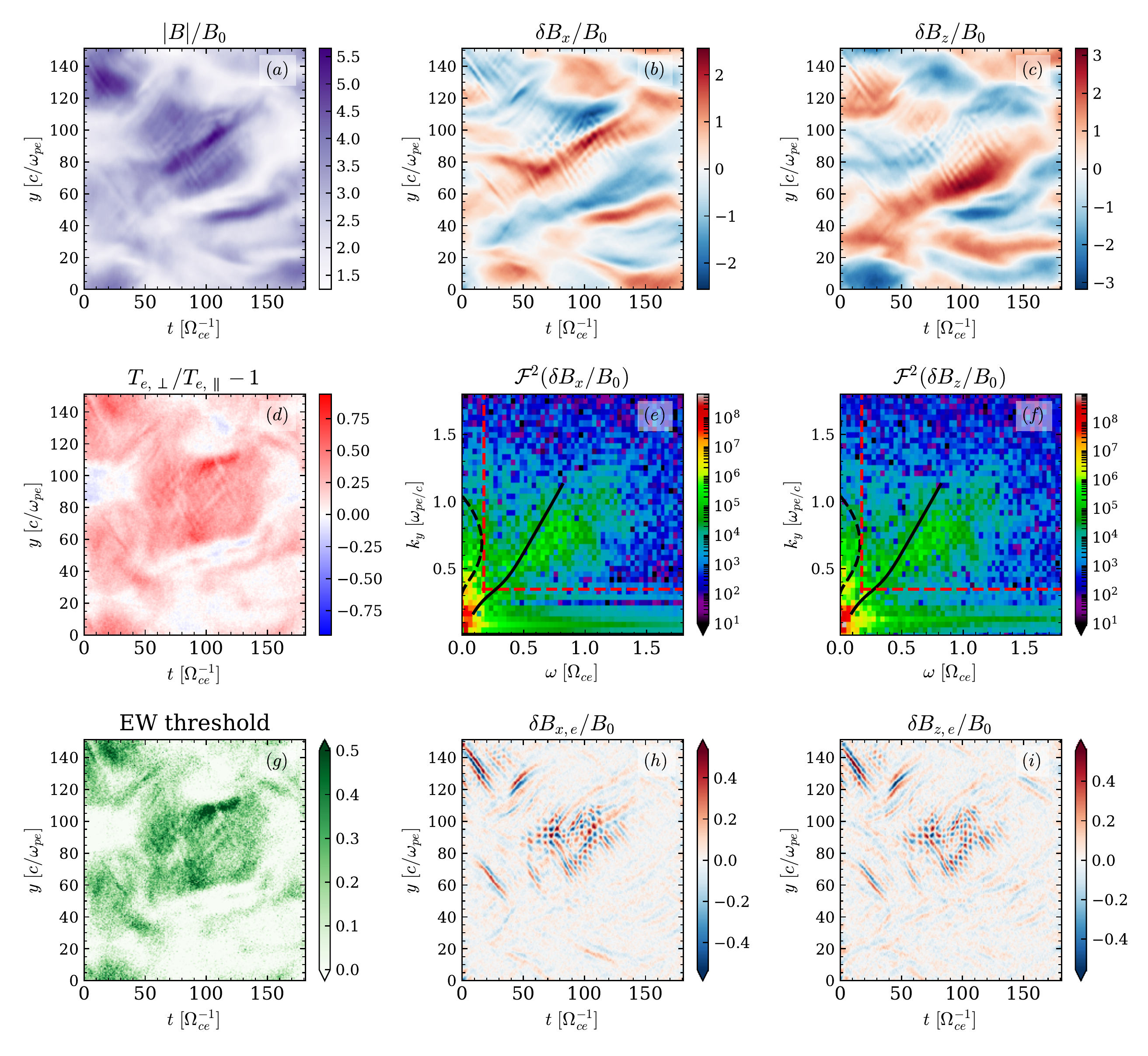}
\end{center}
\caption{Space-time diagrams and power spectra at $\xsh=-122\, c/\omega_{pe}\sim -2.5\rli$ behind the shock (as indicated by the vertical dotted pink lines in \fig{fields} and \fig{elecs}), during the same time interval $24.0\leq\Omega_{ci}t\leq27.4$ as in \fig{cut1}. For panels (a)-(f), see \fig{cut1}, the only difference being that the predictions in panels (e) and (f) (solid black line for the whistler wave frequency, dashed black line for the growth rate) are computed considering the plasma properties only in regions where the electron anisotropy is well above the whistler threshold, more specifically $T_\perpe/T_\pare-1-0.21/\beta_\pare^{0.6}\geq 0.3$. In panel (g), where we indeed plot $T_\perpe/T_\pare-1-0.21/\beta_\pare^{0.6}$, this would correspond to the dark green areas. Since panels (a)-(c) are dominated by long-wavelength slowly-propagating proton modes, we isolate electron waves via a high-pass filter in the power spectra of panels (e) and (f), keeping only the high-$\omega$ high-$k_y$ region delimited by the red dashed lines. The resulting space-time wave pattern is shown in panels (h) and (i), which reveal the presence of electron whistler waves.}\label{fig:cut2}
\end{figure*}

\subsubsection{Electron Whistler Waves}\label{sec:whistler}
The physics of particle irreversible heating that we have described in \sect{phys} relies on two ingredients: a certain level of particle anisotropy, and a mechanism to break the adiabatic invariance. As we have shown above, a large-scale magnetic field amplification (e.g., resulting from shock compression of the upstream field) will lead to electron anisotropy with $T_\perpe>T_\pare$. In turn, this triggers the growth of whistler waves, which scatter the electrons in pitch angle, providing a mechanism to break the adiabatic invariance and generate irreversible heat. Below, we show that the two ingredients needed for entropy increase are indeed present in the two locations where the entropy profile shows the fastest increase (vertical dotted lines in \fig{fields} and \fig{elecs}). 

At the shock (grey dotted line in \fig{fields} and \fig{elecs}), the electron temperatures are driven to $T_\perpe>T_\pare$ by shock-compression of the upstream field, via conservation of the first and second adiabatic invariants. In \fig{cut1}, we show the space-time diagram of various quantities, in the time interval $20.0\leq\Omega_{ci} t\leq27.4$ and along the $y$ extent of the box. The $x$ location is fixed at the shock ramp (more precisely, $\xsh=4 \,c/\omega_{pe}$).
Shock-compression of the upstream field (see Figure \ref{fig:cut1}(a), where $|B|/B_0\sim 2.2$) leads to a temperature anisotropy $T_\perpe/T_\pare-1\simeq0.6$ (Figure \ref{fig:cut1}(d)). Both the field amplification and the degree of temperature anisotropy are nearly constant in time and uniform in $y$. 

As a result of the strong temperature anisotropy, magnetic waves are excited throughout the $y$ range consistently over time. 
Panels (b) and (c) show the space-time diagrams of the magnetic fluctuations $\dbx$ and $\dbz$, revealing the presence of high-frequency and short-wavelength modes (as also seen in \fig{fields}(b) and (c) near the shock). Figure \ref{fig:cut1}(e) and (f) show the corresponding power spectra, as a function of frequency $\omega$ (horizontal axis) and  wavenumber $k_y$ (vertical axis). 
The power spectrum displays a pronounced peak at frequency $\omega\simeq 0.5 \, \Omega_{ce}$ (here $\Omega_{ce}=(m_i/m_e)\Omega_{ci}$ is the electron gyrofrequency) and wavevector $k_y\simeq 0.5 \,\omega_{pe}/c$. We have compared this with linear theory of the electron whistler instability \citep[e.g.][]{Gary1985,Gary1996,Gary2006} by solving the dispersion relation 
\begin{align}
0 & = D^{\pm } (k_y, \Omega)\nonumber\\
& =\Omega^2 - c^2 k_y^2 + \omega_{pi}^2 \frac{\Omega}{k_z v_i}Z(\zeta_i^{\pm})\nonumber\\
& + \omega_{pe}^2\frac{\Omega}{k_z v_{e,\parallel}^2}Z(\zeta_e^{\pm}) 
+ \omega_{pe}^2\left( \frac{T_{e\perp}}{T_{e,\parallel}}-1\right)
\left[1+\zeta_e^{\pm } Z(\zeta_e^{\pm})\right]\,\label{eq:dispersion}
\end{align}
where 
$
\zeta_e^{\pm} = (\Omega \pm \omega_{ce})/k_y v_e,
$
$
v_{e,\parallel} = (2k_{\rm B}T_{e,\parallel}/m_e)^{1/2},
$
$
\zeta_i^{\pm} = (\Omega \pm \omega_{ci})/k_y v_{i,\parallel},
$
$
v_{i} = (2k_{\rm B}T_{i}/m_i)^{1/2}, 
$
and
$Z(\zeta)$ is the plasma dispersion function 
\begin{equation}
Z(\zeta) = \frac{1}{\sqrt{\pi}}\int_{-\infty}^{\infty} dx \frac{\exp(-x^2)}{x-\zeta}.
\end{equation}
The input values of $v_{i}, v_{e,\parallel},{T_{e\perp}}/{T_{e,\parallel}}-1$ for the dispersion relation are taken from the time- and space-averages of the corresponding quantities over the same time period and spatial extent as the space-time diagram in Figure \ref{fig:cut1}.  The resulting theoretical prediction for the real part of the frequency is shown with a black solid line in panels (e) and (f), and it matches extremely well the contours of the power spectrum. The imaginary part of the frequency, i.e., the growth rate of the mode, is plotted with a dashed  black curve. The value of $k_y$ giving the fastest growth agrees well with the location of the peak of the power spectrum ($k_y\simeq 0.5 \,\omega_{pe}/c$). The excellent agreement between the simulation data and the electron whistler  dispersion  relation confirms that the waves in the shock ramp are produced by the electron whistler instability.

Figure \ref{fig:cut2} shows similar plots at the location indicated in \fig{fields} and \fig{elecs} with a vertical dotted pink line, at $\xsh\simeq-2.5\rli\simeq -122\,c/\omega_{pe}$. Here, field amplification is driven by a combination of two effects: the density (and so, the frozen-in magnetic field) experiences another large-scale compression; in addition, the proton-driven waves shown in \fig{fields} further increase the local magnetic field intensity. 

As compared to \fig{cut1}, the space-time diagrams show now a higher degree of inhomogeneity, imprinted by the anisotropy-driven long-wavelength proton modes. These fluctuations co-exist with weaker small-wavelength high-frequency modes, which only appear in localized patches (e.g., at $x\sim 80\compe$ and $t\sim 100\,\Omega_{ce}^{-1}$ in \fig{cut2}(b) and (c)). The high-frequency waves are generated in regions where field amplification (\fig{cut2}(a) at $x\sim 80\compe$ and $t\sim 100\,\Omega_{ce}^{-1}$) causes the electron anisotropy (\fig{cut2}(d)) to exceed the threshold for whistler growth (\fig{cut2}(g)), which is given by
\be\label{eq:threshwhis}
\frac{T_\perpe}{T_\pare}-1\simeq \frac{0.21}{\beta_\pare^{0.6}}~,
\ee
where $\beta_\pare=8 \pi n_e k_{\rm B} T_\pare /B^2$ is the local value of the parallel electron beta \cite{Gary2005}. 

Figure \ref{fig:cut2}(e) and (f) show the power spectra of $\delta B_x$ and $\delta B_z$. Most of the power is concentrated in low-frequency long-wavelength modes, generated by the proton cyclotron or mirror instabilities. However, there is still an appreciable amount of power in high-frequency short-wavelength modes peaking at $\omega\sim 0.7\,\Omega_{ce}$ and $ k_y \sim 0.7\, \omega_{pe}/c$. 
We apply a high-frequency short-wavelength filter, in order to isolate the top right region in \fig{cut2}(e) and (f) (the cutoff frequency and wavenumber of our filter are shown with dashed red lines). This allows us to extract (via an inverse Fourier transform) the space-time wave patterns of  high-frequency short-wavelength modes, which are shown in panels (h) and (i). The two panels confirm that short-wavelength modes exist only in regions where the electron temperature anisotropy exceeds the electron whistler threshold (Figure \ref{fig:cut2}(g) at $x\sim 80\compe$ and $t\sim 100\,\Omega_{ce}^{-1}$). We have measured the average electron and proton temperatures and densities in the region where the whistler threshold is appreciably exceeded ($T_\perpe/T_\pare-1-0.21/\beta_\pare^{0.6}\geq 0.3$), in order to obtain linear theory predictions. The real part and imaginary part of the resulting dispersion relation are plotted in \fig{cut2}(e) and (f) with solid and dashed black lines, respectively. The good agreement with the power spectra extracted from our shock simulation confirms the presence of patches of whistler waves in the second ramp (at $\xsh\sim -2.5\rli$) of the electron entropy profile.

To summarize, we have identified two major sites of electron entropy production in the shock downstream. One is at the shock ramp, and the other is at a distance of $\sim 2.5\rli$ behind the shock, where density compression and proton-driven waves both contribute to magnetic field amplification. 
Both sites show the presence of electron whistler waves triggered by electron temperature anisotropy. Whistler waves provide the pitch-angle scattering required to break electron adiabatic invariance and to generate entropy. In the following two sections, we further elucidate the physics of entropy production in these two sites, by means of periodic box simulations.

\section{Electron Heating in the Shock Ramp}\label{sec:ramp}
The first increase in electron entropy happens in the shock ramp. As a result of the shock-compression of the upstream field ($B\propto n$ by flux freezing), electrons become anisotropic and they trigger whistler waves. Below, we model the shock compression in a periodic box using a novel form of the PIC equations introduced in \citet{Sironi2015,Sironi2015a}, which incorporates the effect of a large-scale compression of the system. We briefly describe the simulation setup in \sect{setup1}, we discuss periodic box simulations applicable to our reference shock run in \sect{ref1}, and we describe the dependence on mass ratio in \sect{mass1}.

\begin{figure}[tbh]
\begin{center}
\includegraphics[width=0.5\textwidth]{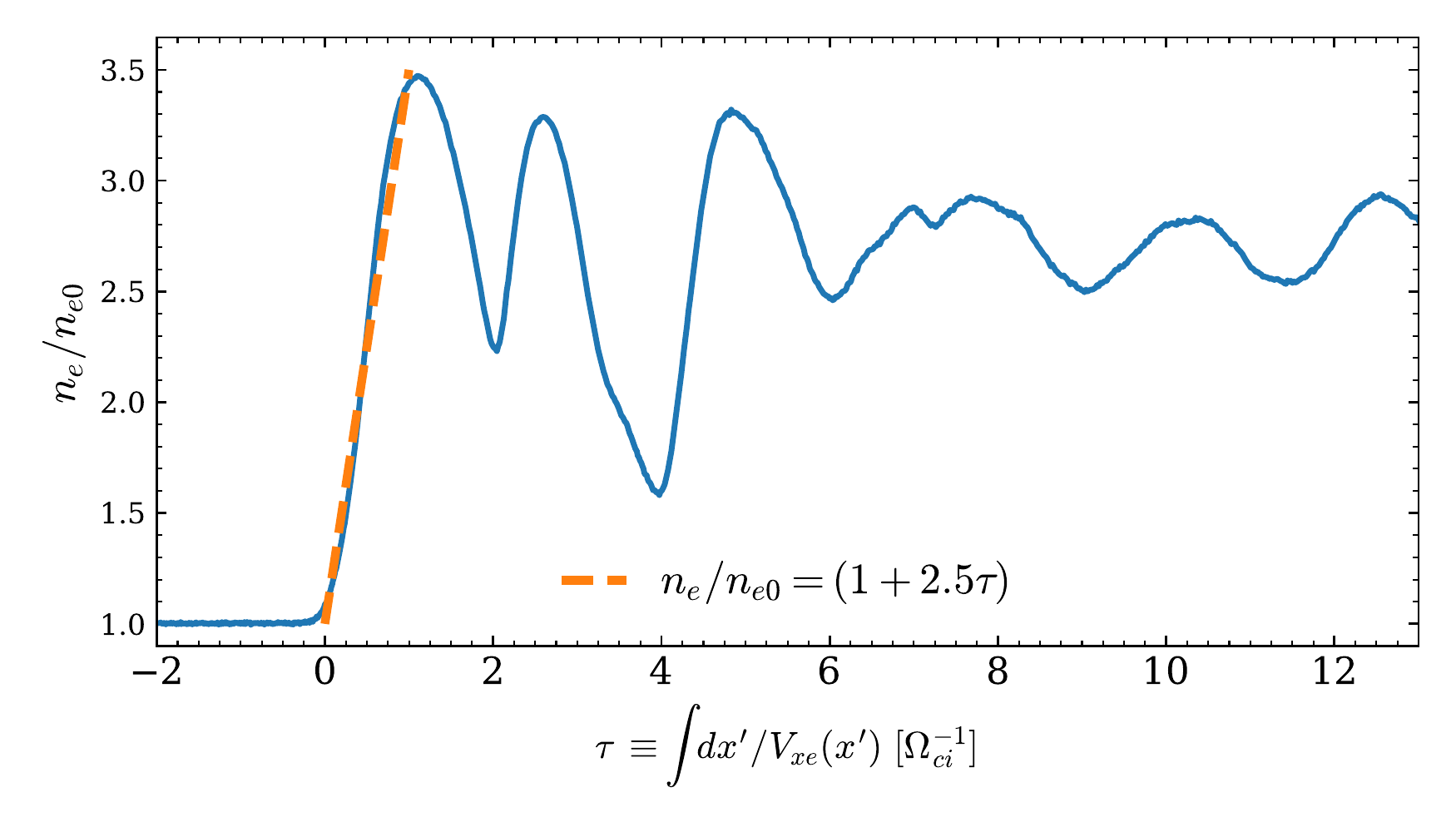}
\end{center}
\caption{As a function of the comoving time of the electron fluid defined in \eq{electime}, we present the density profile experienced by electrons as they propagate from upstream to downstream (solid blue line). The time axis is shifted such that $\tau=0$ just ahead of the shock. The shock-compression felt by incoming electrons can be approximated as $n_e/n_{e0}=(1+q\,t)$with compression rate $q=2.5\,\Omega_{ci}$ (orange dashed line).}\label{fig:compress_rate}
\end{figure}
\subsection{Simulation Setup}\label{sec:setup1}
To emulate the conditions for electrons in the shock ramp, we set up a suite of compressing box experiments, using the method introduced in \citet{Sironi2015,Sironi2015a}.  Here, we report only its main properties. 
We  solve Maxwell's equations and the Lorentz force in the fluid {\it comoving} frame, which is related to the {\it laboratory} frame by a Lorentz boost. In the comoving frame, we define two sets of spatial coordinates, with the same time coordinate. The {\it unprimed} coordinate system has a basis of unit vectors, so it is the appropriate coordinate set to measure all physical quantities. Yet, it is convenient to re-define the unit length of the spatial axes in the comoving frame such that a particle subject only to compression stays at fixed coordinates. This will be our {\it primed} coordinate system. Then, compression with rate $q$ is accounted for by the diagonal matrix 
\be\label{eq:eqL}
\L=\frac{\partial \bmath{x}}{\partial \bmath{x}'}=
\left(
\begin{array}{ccc}
\qt^{-1} \,\,\,& 0 \,& 0 \\
 0\,\,\,& 1 \,& 0 \\
 0 \,\,\,& 0 \,&1\\
\end{array}\right)~~~,
\ee
which has been tailored for compression along the $x$ axis, as expected in our shock. 

A uniform ordered magnetic field $\bmath{B}_0$ is initialized along the $y$ direction (in analogy to the shock setup). We define $\Omega_{ci}$ as the proton Larmor frequency in the initial field $B_0$.
Maxwell's equations in the primed coordinate system \citep{Sironi2015} prescribe that the field will grow in time as $\bvec_0\qt$, which is consistent with flux freezing (the particle density in the box increases at the same rate). From the Lorentz force in the compressing box \citep{Sironi2015}, the component of particle momentum aligned with the field does not change during compression, whereas the perpendicular momentum increases as $\propto\sqrt{1+q\,t}$. This is consistent with the conservation of the first and second adiabatic invariants.

This method is implemented for 1D, 2D and 3D computational domains, with periodic boundary conditions in all directions. 
In the previous section, we have shown that the whistler instability is the dominant mode in the shock ramp. Its wavevector is nearly aligned with the field direction (i.e., along $\hat{y}$)
It follows that the evolution of the dominant mode can be conveniently captured by means of 1D simulations with the computational box oriented along $y$, which we will be employing in this section. Yet, all three components of electromagnetic fields and particle velocities are tracked. In 1D simulations, we can employ a large number of particles per cell (typically, $10^4$ per cell) so we have adequate statistics for the calculation of the electron specific entropy from the phase space distribution function. In addition, in 1D simulations we can readily extend our results up to the realistic mass ratio. Even though we only show results from 1D runs, we have checked that the main conclusions hold in 2D. 

As a result of the large-scale compression encoded in \eq{eqL}, both electrons and protons will develop a temperature anisotropy, and we should witness the development of both electron and proton anisotropy-driven modes. However, in our reference shock, no proton modes grow in the shock ramp (they only develop a few Larmor radii behind the shock). For this reason, in our compressing box runs, we artificially inhibit the update of the proton momentum (effectively, this corresponds to the case of infinitely massive protons, which only serve as  a charge-neutralizing fluid).

The compression rate $q$ is measured directly from our reference shock simulation. There, we can quantify the profile of electron density as a function of the co-moving time of the electron fluid, which follows from
\begin{equation}\label{eq:electime}
\tau \equiv \int \frac {dx'}{V_{xe}(x')}
\end{equation}
where $V_{xe}$ is the electron fluid velocity in the shock  frame, and the integral goes from the upstream to the downstream region. 
Figure \ref{fig:compress_rate} shows the density profile as a function of $\tau$ from our reference run. The density oscillates on a timescale comparable to the proton gyration time $\Omega_{ci}^{-1}$, which is expected given that the proton dynamics controls the shock structure. 
At the ramp starting near $\tau = 0$, the electron density increases by a factor of $\sim 3.5$ within $\sim1\, \Omega_{ci}^{-1}$. Even though the density increase is not perfectly linear, we find that a linear approximation with $q=2.5\,\Omega_{ci}$ provides a reasonable fit (see the orange dashed line in \fig{compress_rate}).
We remind that our electron heating model is agnostic of the exact profile of density compression, as long as the compression rate and the resulting field amplification rate are much slower than the electron gyration frequency. 

Below, we fix $q=2.5\,\Omega_{ci}$. With increasing mass ratio, the separation between $q$ and the electron cyclotron frequency $\Omega_{ce}$ will increase as $m_i/m_e$. As in our reference shock run, electrons are initialized to have $k_{\rm B}T_{e0} = 10^{-2}m_e c^2$, and the strength of the background magnetic field is set so that $\beta_{p0}= 16$. We resolve the electron skin depth with 10 cells, so the Debye length is marginally resolved. The box extent along the $y$ direction is fixed at $43\ c/\omega_{pe}$, which is sufficient to capture several wavelengths of the electron whistler  instability. The box size does not need to scale with the mass ratio, since we are artificially excluding the proton physics.

\begin{figure}[tbh]
\begin{center}
\includegraphics[width=0.355\textwidth]{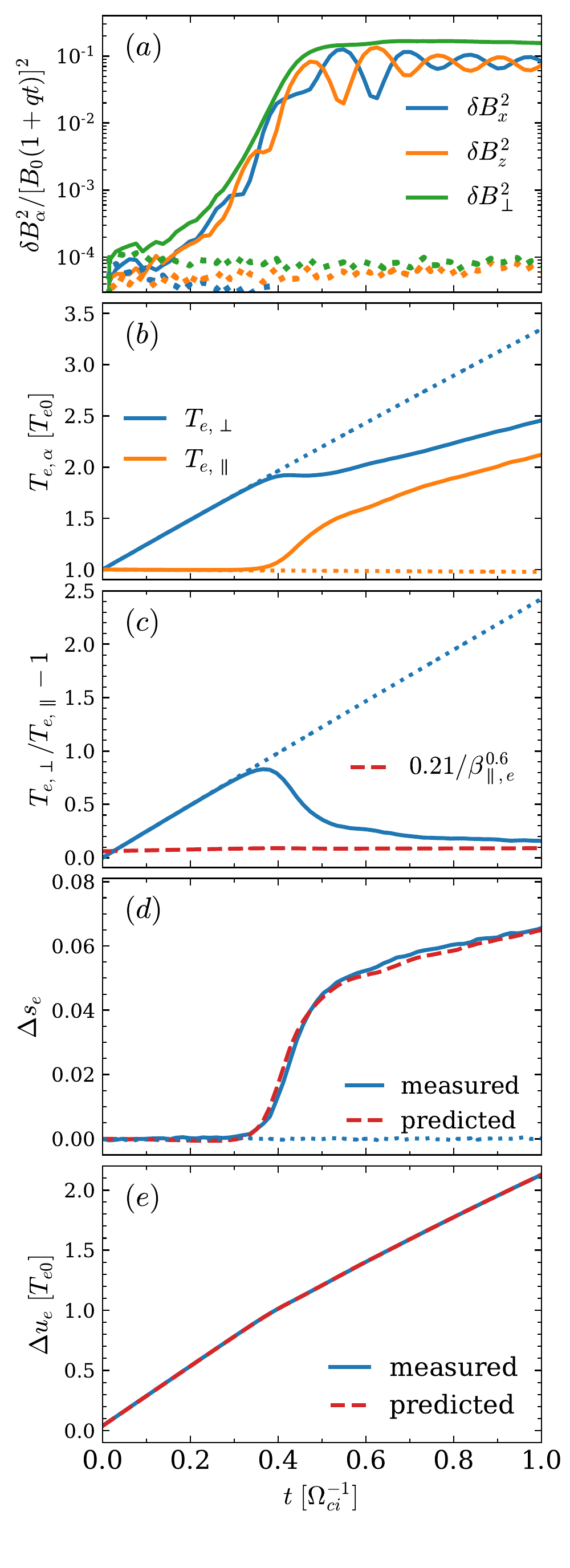}
\end{center}
\caption{Time evolution of various space-averaged quantities in a 1D periodic box whose compression rate $q=2.5\,\Omega_{ci}$ is chosen to mimic the effect of the shock ramp. We compare two field geometries, with background field lying either along the $y$ axis of the simulation box (``in-plane'' configuration, solid lines) or along the $z$ direction perpendicular to the box (``out-of-plane'' configuration, dotted lines): (a) energy in magnetic field fluctuations, normalized to the energy of the compressed magnetic field (the legend is appropriate for the in-plane configuration, whereas for the out-of-plane case the orange line refers to $\dby^2$); (b) electron temperature perpendicular ($T_\perpe$, blue lines) and parallel ($T_\pare$, orange lines) to the background field; (c) electron temperature anisotropy (blue lines), and comparison with the threshold of the electron whistler instability, as in \eq{threshwhis} (dashed red line); (d) electron entropy change, measured from the electron distribution function as in \eq{sdef} (blue solid) or predicted from \eq{dselece} (red dashed); (e) electron energy increase in units of $k_{\rm B}T_{e0}$, measured directly (blue solid) or predicted using \eq{due} (red dashed).}\label{fig:boxinoutplane}
\end{figure}

\subsection{Application to the Reference Shock}\label{sec:ref1}
As in our reference shock run, we employ a reduced mass ratio $m_i/m_e=49$. In the periodic compressing box, this means that our choice of $q=2.5\,\Omega_{ci}$ leads to a compression rate that is a factor of $\sim20$ lower than the electron gyration frequency. 

To highlight the importance of the electron whistler instability in facilitating electron entropy production, in Figure \ref{fig:boxinoutplane} we compare two simulations, one with the background field ${\b B}_0$  in the $z$ direction, the other one with ${\b B}_0$ along the $y$ direction.
Since our simulation box is oriented along $y$ and the dominant wavevector of the electron whistler  instability is parallel to the background field, if the field lies along $z$ (which we shall call ``out-of-plane'' case, and indicate with dotted lines) we artificially suppress the growth of electron whistlers. By comparing it with the ``in-plane'' simulation with the field along $y$ (solid lines), which does allow for whistler wave growth, we can demonstrate the importance of the electron whistler instability for entropy production.

In the absence of electron-scale instabilities that would break the adiabatic invariance, the out-of-plane simulation is expected to follow adiabatic scalings. In fact, in the out-of-plane simulation, we see that $\Teperp\propto B\propto (1+qt)$ (blue dotted line in \fig{boxinoutplane}(b)), while $T_{e,\parallel}\propto (n/B)^2\propto \;$const (orange dotted line in \fig{boxinoutplane}(b)), as expected from the double adiabatic theory. Since no whistler waves grow (notice that the fields stay at the noise level, see the dotted lines of \fig{boxinoutplane}(a)), no mechanism exists that can transfer heat from the perpendicular to the parallel temperature, and the electron entropy remains constant.

The in-plane simulation shows a different behavior. Initially, $\Teperp$ and $T_{e,\parallel}$ follow the double adiabatic trends (solid lines in \fig{boxinoutplane}(b)). At $\Omega_{ci}t\sim 0.3$, the increasing temperature anisotropy (blue solid line in panel (c)) leads to the exponential growth of electron whistler waves (solid lines in \fig{boxinoutplane}(a)). At $\Omega_{ci}t\sim 0.4$, the wave energy is strong enough to pitch-angle scatter the electrons. As a result, heat is transferred from the perpendicular to the parallel direction. Both $\Teperp$ and $T_{e,\parallel}$ deviate from the adiabatic scalings and the temperature anisotropy is reduced.

At $t\sim 0.4\, \Omega_{ci}^{-1}$, with the onset of pitch-angle scattering and the consequent breaking of adiabatic invariance, the electron specific entropy starts to rise (solid blue line in Figure \ref{fig:boxinoutplane}(d)). 
The most rapid entropy increase happens near the end of the exponential whistler growth, at $t\sim 0.4 - 0.5\, \Omega_{ci}^{-1}$. Here, the electron anisotropy is still large, and at the same time whistler waves are sufficiently powerful to provide effective pitch-angle scattering. In other words, both terms in the square brackets of either \eq{dselec} or \eq{dsperp} are large.
After the exponential growth, the electron whistler waves enter a secular phase where the wave energy (normalized  to the compressed background  field energy) stays almost constant (solid green line in Figure \ref{fig:boxinoutplane}(a)). In this phase, whistler waves are continuously generated as the large-scale compression steadily pushes the electron anisotropy slighly above the threshold of marginal stability (indicated by the red dashed line in \fig{boxinoutplane}(c)). Both the ingredients needed for entropy increase (i.e., nonzero electron anisotropy and efficient pitch angle scattering mediated by whistler waves) persist during the secular phase, leading to further increase in the electron entropy.

In \fig{boxinoutplane}, we also explicitly validate the heating model described in \sect{phys}. Following \eq{energyconserv}, the electron energy per particle should change as 
\be\label{eq:enel2}
du_e&=&k_{\rm B}T_{\perpe}d\ln B+k_{\rm B}T_{e,\parallel}d\ln\left(\frac{n}{B}\right)-de_{w,e}\nonumber\\
&=&k_{\rm B}T_{\perpe}d\ln B-de_{w,e}\,\label{eq:due}
\ee
where $e_{w,e}$ is the energy per particle in whistler waves (as we have discussed in \sect{phys}, the electron energy transferred to electron modes stays entirely in the waves, so $de_{w,e}=de_{w \, {\rm tot},e}$), and we have used the fact that $n/B\propto\;$const. In Figure \ref{fig:boxinoutplane}(e), the blue solid line shows the measured change of electron internal energy from the in-plane run, while the red dashed line is obtained by integrating \eq{enel2}. We find excellent agreement between simulation results and our electron  heating model. 

The validation can also be extended to the entropy measurement, as we do in \fig{boxinoutplane}(d). Again, the blue solid line shows the measured change in electron specific entropy (computed from the distribution function as in \eq{sdef}), while the red dashed line is obtained by integrating
\be
ds_e 
& = &\left(\frac{1}{2}d\ln {T_{e,\parallel}}\right)\left(1-\frac{T_{e,\parallel}}{T_{\perpe}}\right)-\frac{d e_{w,e}}{T_{\perpe}}~,\label{eq:dselece}
\ee
which follows from Equation \eqref{eq:dselec} 
(an equivalent form can be obtained from \eq{dsperp}). Once again, the model matches the simulation results extremely well.

\begin{figure}[tbh]
\begin{center}
\includegraphics[width=0.4\textwidth]{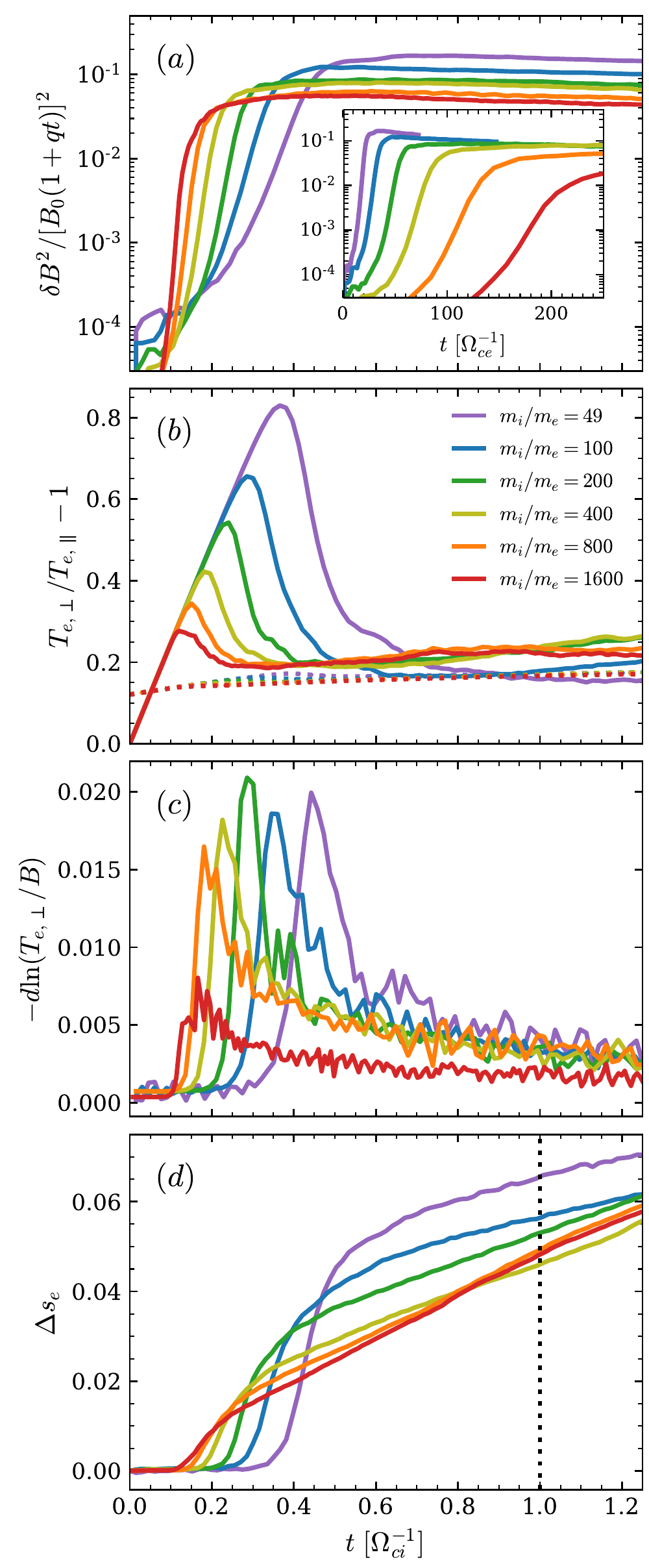}
\end{center}
\caption{Dependence on mass ratio  (up to $m_i/m_e=1600$) of various space-averaged quantities in a 1D periodic box with compression rate $q=2.5\,\Omega_{ci}$ (the legend is in panel (b)). The background field is aligned with the box (in-plane configuration). We plot: (a) energy in magnetic field fluctuations, normalized to the energy of the compressed  field; (b) electron temperature anisotropy (solid lines) and threshold condition for the electron whistler instability (dotted lines with the same color coding as the solid lines); (c) rate of violation of adiabatic invariance $-d\ln (T_\perpe/B)$; (d) electron entropy change, measured from the electron distribution function as in \eq{sdef}. After $\Omega_{ci} t\sim 1$ (vertical dotted black line in panel (d)), which corresponds to the end of the compression phase in the shock ramp, the entropy change is nearly independent of the mass ratio.}\label{fig:compressbox_massratio}
\end{figure}

\subsection{Dependence on the Mass Ratio}\label{sec:mass1}
We now extend our compressing box experiments up to the realistic mass ratio and show that the electron entropy increase is nearly insensitive to $m_i/m_e$ (as long as the mass ratio is larger than a few tens).
Figure \ref{fig:compressbox_massratio} compares the evolution of the whistler wave energy (panel (a)), the electron temperature anisotropy (panel (b)), the rate $-d\ln (T_\perpe/B)$ of breaking adiabatic invariance (panel (c)) and the electron entropy increase (panel (d)) when varying the mass ratio from $m_i/m_e=49$ up to $m_i/m_e=1600$ (from purple to red, see the legend in the second panel). Since we fix the compression rate to be $q=2.5\,\Omega_{ci}$, a larger mass ratio corresponds to a lower compression rate in units of the electron gyration frequency $\Omega_{ce}=(m_i/m_e)\Omega_{ci}$.

Initially, the electron temperature anisotropy grows linearly in time as $T_\perpe/T_\pare-1= qt$, as a result of the large-scale compression. This proceeds until the energy in whistler waves reaches a fraction $\sim 3\times10^{-2}$ of the compressed background field energy (\fig{compressbox_massratio}(a)). At this point, whistler waves are sufficiently strong to scatter the electrons in pitch angle, breaking their adiabatic invariance and reducing the electron anisotropy by transferring energy from the perpendicular to the parallel component. In fact, notice that the peak in panel (c), i.e., the time when the electron adiabatic invariance is most violently broken, always corresponds to the time when the electron anisotropy in panel (b) shows the sharpest decrease.

The onset of efficient pitch-angle scattering (and so, the peak time of electron anisotropy) occurs earlier at higher mass ratio, at a time that decreases from $t\sim 0.35\, \Omega_{ci}^{-1}$ at $m_i/m_e = 49$ down to $t\sim 0.1\, \Omega_{ci}^{-1}$ at $m_i/m_e = 1600$. This can be understood from the competition between the large-scale compression rate (which increases the electron anisotropy) and the growth rate of whistler waves (which try to reduce the anisotropy via pitch angle scattering). The compression rate in units of the electron cyclotron frequency is  $q=2.5 \,(m_e/m_i)\Omega_{ce}$, while the whistler growth rate (also in units of $\Omega_{ce}$) depends on how much the anisotropy exceeds the whistler threshold in \eq{threshwhis}. In order to balance the two rates, a higher anisotropy is needed for larger $m_e/m_i$, i.e., for lower mass ratios. This has two consequences: first, the growth rate of the whistler instability (normalized to $\Omega_{ce}$) will  decrease at higher $m_i/m_e$, as indeed confirmed by the inset of \fig{compressbox_massratio}(a); second, lower peak anisotropies (and so, earlier onsets of efficient pitch angle-scattering) will be achieved at higher mass ratios, which explains the trend seen in \fig{compressbox_massratio}(b). In addition, since the energy of whistler waves ultimately comes from the free energy in electron anisotropy, higher mass ratios display weaker levels of whistler wave activity (panel (a)). 

The electron entropy evolution in \fig{compressbox_massratio}(d) can be separated into two stages. In the first phase (which, for $m_i/m_e=49$, occurs at $t\sim 0.45\, \Omega_{ci}^{-1}$), the electron entropy grows quickly. This stage corresponds to the late exponential phase of whistler wave growth (and so, we shall call it ``exponential phase''), when both the electron anisotropy (panel (b)) and the rate of breaking adiabatic invariance (panel (c)) --- i.e., the two ingredients needed for efficient entropy production --- are large. Since higher mass ratios reach lower levels of electron anisotropy, the entropy produced during this stage  is a decreasing function of $m_i/m_e$, as seen in \fig{compressbox_massratio}(d) (compare the purple line growth around $t\sim 0.45\, \Omega_{ci}^{-1}$ with the red line around $t\sim 0.15\, \Omega_{ci}^{-1}$). 
After whistler waves have reached saturation, the electron entropy still increases, in a phase which we shall call ``secular''. Here, the electron anisotropy stays close to the threshold of marginal stability (indicated in \fig{compressbox_massratio}(c) by the dotted lines, with the same color coding as the solid curves). Continuous pitch-angle scattering (and so, persistent violation of adiabatic invariance) is needed to oppose the steadily-driven compression and maintain the system close to marginal stability. It is then expected that entropy will continuously increase during the secular phase, albeit at a lower rate than in the exponential stage. For $m_i/m_e\gtrsim400$, the electron anisotropy at late times is nearly insensitive to $m_i/m_e$ (compare yellow, orange and red lines at $\Omega_{ci}t\gtrsim0.4$ in panel (b)), which explains why the entropy growth in the secular phase is nearly the same for all $m_i/m_e\gtrsim400$ (\fig{compressbox_massratio}(d)).

From \fig{compressbox_massratio}(d), we can infer how the entropy increase in the shock ramp should scale with mass ratio. Since the compression in the shock ramp lasts about one proton gyration time, we compare the entropy curves at $\Omega_{ci}t\sim1$, as indicated by the vertical dotted black line in panel (d). When the  mass ratio increases from $m_i/m_e=49$ to $m_i/m_e=1600$ (i.e., more than a factor of $32$),  the entropy produced until  $\Omega_{ci}t=1$ decreases from $0.065$ to $0.048$, only a $\sim 30\%$ decrease. The dependence on mass ratio would be far more pronounced if we were only to consider the entropy produced during the exponential phase. However, higher mass ratio runs have earlier onset times, as we have explained above, so they spend more time (within the first $\Omega_{ci}^{-1}$) in the secular phase, as compared to lower mass ratios. In summary, most of the entropy production at lower mass ratios happens during the exponential phase, whereas at higher mass ratios the secular phase lasts longer and thus compensates for the lower level of entropy generated during the exponential stage. The net effect is that the entropy increase in our compressing box with $m_i/m_e=1600$ is only slightly smaller than for  $m_i/m_e=49$. The same conclusion should hold also for our reference shock.

\section{Electron Heating by Proton-Driven Waves}\label{sec:waves}
In the downstream region of our reference shock, at a distance of $\sim 2.5\rli$ from the shock front, the electron entropy shows a second phase of rapid increase. Here, a large-scale density compression co-exists with the growth of proton-driven waves, and both contribute to magnetic field amplification and irreversible electron heating. The new concept here is the effect of proton-driven waves, so we focus on that in this section. We demonstrate that magnetic fluctuations induced by a proton temperature anisotropy can naturally lead to an increase in electron entropy, {\it even in the absence of a large-scale compression}. We employ periodic simulation domains with the standard form of the PIC equations (as opposed to what we used in the previous section) and set up a population of anisotropic protons, with a degree of anisotropy  motivated by our reference shock run.
We discuss the simulation setup in \sect{setup2}, we discuss periodic box simulations applicable to our reference shock run in \sect{ref2}, and we describe the dependence on mass ratio in \sect{mass2}.

\subsection{Simulation setup}\label{sec:setup2}
In order to study the role of anisotropy-driven proton modes in producing electron irreversible heating, we set up a periodic simulation box with anisotropic protons. The simulation is initialized to approximate the conditions right after the shock transition. The protons are initialized as a bi-Maxwellian distribution with $T_{i0,\perp}/T_{i0,\parallel}\sim 7$, as observed just behind the shock in \fig{protons}(g). The value of $T_{i0,\parallel}$ is the same as in the shock upstream (in fact, the parallel proton temperature is nearly uniform across the shock, see the orange line in \fig{protons}(f)).
The electron temperature increases  roughly by a factor of two across the shock (\fig{elecs}(e)), so the electrons in the tests here are initialized with  $T_{e0}\sim 2\times 10^{-2}m_e c^2/k_{\rm B}$ (note that in the shock upstream the electron temperature was $10^{-2}m_e c^2/k_{\rm B}$). We take electrons to be isotropic, since the fast growth of whistler waves in the shock ramp ensures that the degree of downstream electron anisotropy is low (see the post-shock region in \fig{elecs}(f)). We also take into account that both density and magnetic field strength have increased by a factor of $\sim 2.5$ as compared to the shock upstream (\fig{protons}(a)).

We resolve the electron  skin depth with $7$ cells in order to (marginally) capture the electron Debye length. Since both the proton cyclotron instability, which dominates over the mirror mode in the downstream of our reference run, and the electron whistler instability 
have the fastest growing wavevector aligned with the background field, we  employ 1D simulation domains with the box aligned with the  $y$ direction of the background magnetic field. Thanks to the reduced dimensionality of our computational domain, we can employ a large number of particles per cell ($10^4$). Therefore, we have adequate statistics for the calculation of the electron specific entropy  and we can properly control the effect of numerical heating. In addition, in 1D simulations we can readily extend our results up to the realistic mass ratio. 
The length of the computational box is $1512$ cells for $m_i/m_e=49$. Since the fastest growing mode of the proton cyclotron instability has a wavevector $\sim \omega_{pi}/c$, we increase the number of cells in our computational domain proportional to $\propto \sqrt{m_i/m_e}$, to include the same number of proton skin depths (and so, the same number of proton cyclotron wavelengths).\footnote{Beyond $m_i/m_e=400$, we also increase the number of computational particles per cell proportional to $\sqrt{m_i/m_e}$, in order to minimize numerical heating effects.}

\begin{figure*}[tbh]
\begin{center}
\includegraphics[width=0.65\textwidth]{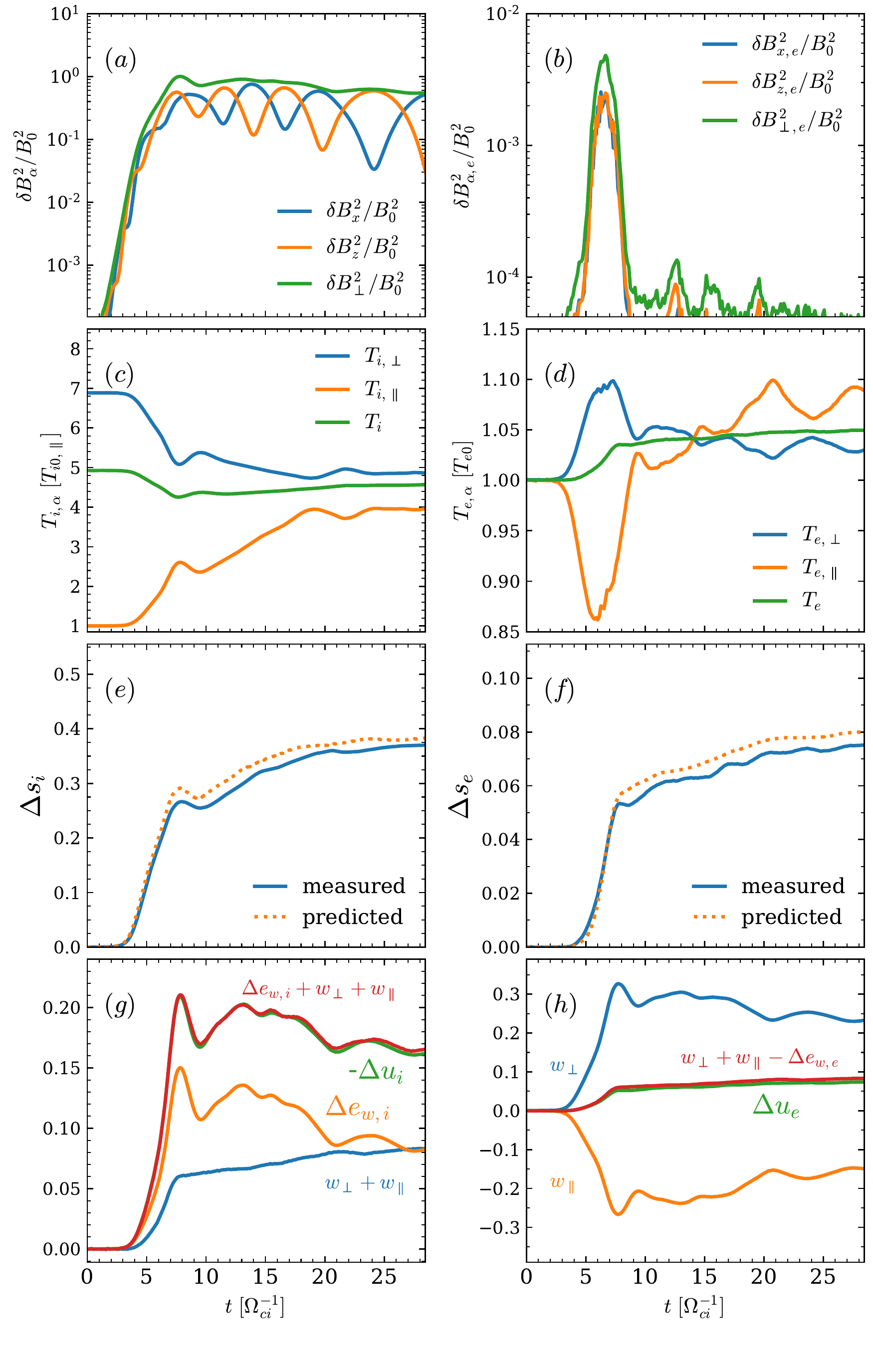}
\end{center}
\caption{Time evolution of various space-averaged quantities in a 1D periodic box initialized with anisotropic protons, to mimic the shock conditions in the downstream. The background field is aligned with the box (in-plane configuration).  We plot: (a) total energy in magnetic field fluctuations, normalized to the energy of the initial field; (b) energy in electron-scale fluctuations, extracted using the high-pass filter in frequency and wavenumber indicated by the red dashed lines in the power spectra of \fig{undriven2Dplot}(e) and (f); (c) proton and (d) electron temperature perpendicular (blue lines) and parallel (orange lines) to the background field, together with the mean temperature (green lines); (e) proton entropy change, measured from the proton distribution function or predicted from \eq{prot1} (orange dotted);  (f) electron entropy change, measured from the electron distribution function (blue solid) or predicted from \eq{dseions} (orange dotted); (g) proton energy change in units of $k_{\rm B}T_{i0}$, measured directly (green) or predicted using \eq{proten} (red); (h) electron energy increase in units of $k_{\rm B}T_{e0}$, measured directly (green) or predicted using \eq{dueions} (red). For other curves in panels (g) and (h), see the text.}\label{fig:undriven1Dplot}
\end{figure*}

\begin{figure*}[tbh]
\begin{center}
\includegraphics[width=0.85\textwidth]{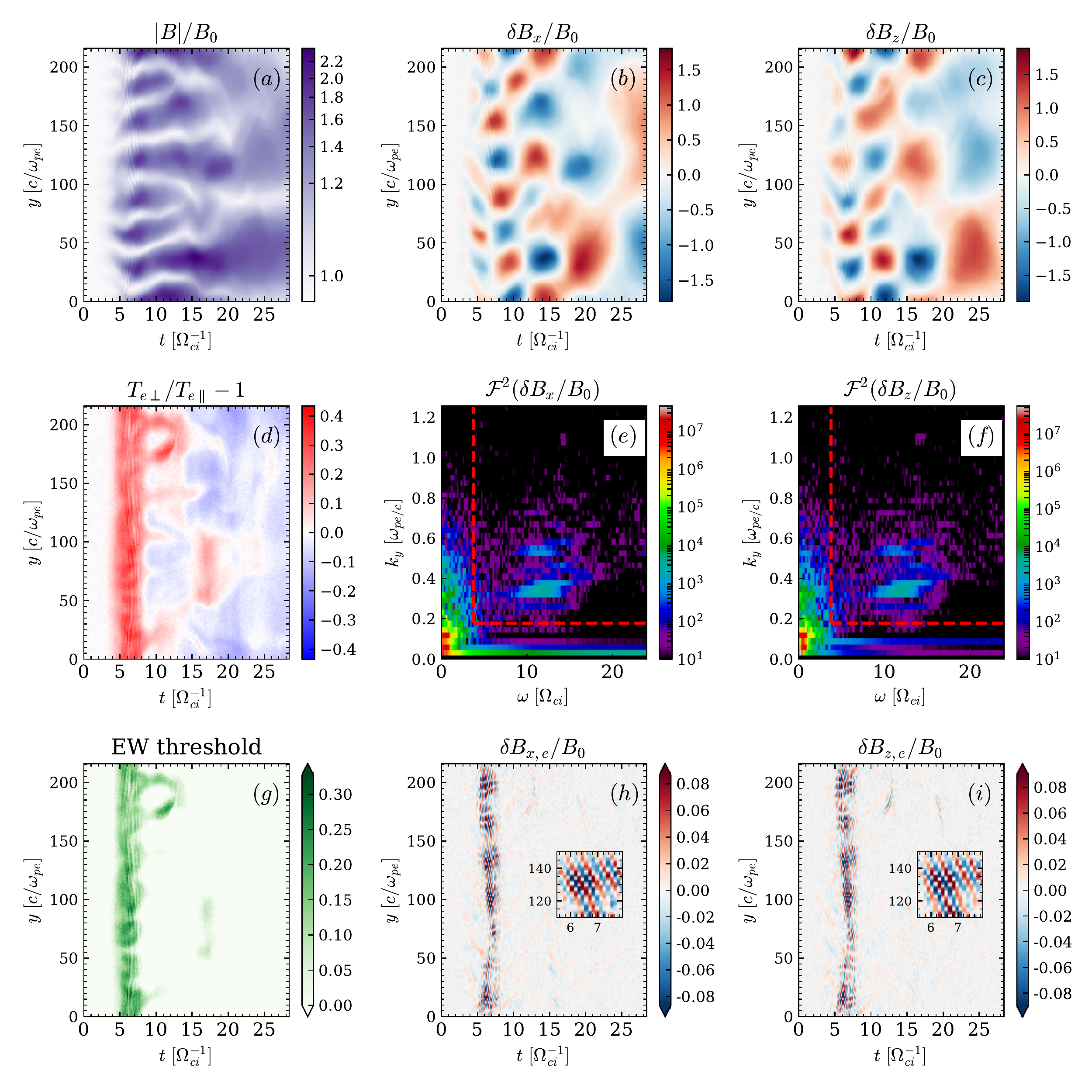}
\end{center}
\caption{Space-time diagrams and power spectra of a 1D periodic box initialized with anisotropic protons. The panels are the same as in \fig{cut2}, with the only difference that the time unit here is $\Omega_{ci}^{-1}$ (and frequencies are normalized to $\Omega_{ci}$).
As in \fig{cut2}, since panels (a)-(c) are dominated by long-wavelength slowly-propagating proton modes, we isolate electron waves via a high-pass filter in the power spectra of panels (e) and (f), keeping only the high-$\omega$ high-$k_y$ region delimited by the red dashed lines. The resulting space-time wave pattern is shown in panels (h) and (i), whose insets clearly reveal the presence of electron whistler waves.}\label{fig:undriven2Dplot}
\end{figure*}

\subsection{Application to the Reference Shock}\label{sec:ref2}
In order to compare the results obtained from the periodic box simulations with the reference shock run, in Figure \ref{fig:undriven1Dplot} and \ref{fig:undriven2Dplot} we show the evolution of our periodic system for $m_i/m_e=49$. As a result of the initial proton temperature anisotropy, the proton cyclotron instability develops, generating exponentially-growing waves with $\dbx$ and $\dbz$ components (\fig{undriven1Dplot}(a)). As shown in Figure \ref{fig:undriven2Dplot}(b) and (c), the growing waves are dominated by modes with wavelength at the proton inertial scale (for $m_i/m_e=49$, the proton skin depth is $c/\omega_{pi}=7\, c/\omega_{pe}$) and frequency comparable to the proton gyration frequency, as expected for the proton cyclotron instability.

At $t\sim 4\,\Omega_{ci}^{-1}$, when the energy in proton cyclotron waves reaches a fraction $\sim 10^{-1}$ of the background field energy, efficient pitch-angle scattering quickly reduces the proton temperature anisotropy (Figure \ref{fig:undriven1Dplot}(c)). During the isotropization process, the proton specific entropy increases (Figure \ref{fig:undriven1Dplot}(e) at $t\sim 5\,\Omega_{ci}^{-1}$). 

The heating model described in \sect{phys} can be applied to protons, keeping in mind that  in the current setup no perturbations in density or magnetic field exist on scales larger than the proton scales (so, no work is being done on the protons). It follows that the perpendicular and parallel energy per proton change as 
\begin{equation}
d u_{i,\perp}=-d q_{i,\perp\to\parallel}-d e_{w,i\perp}\,\label{eq:duiperp}
\end{equation}
\begin{equation}
d u_{i,\parallel}=d q_{i,\perp\to\parallel}-d e_{w,i\parallel}~,\label{eq:duipara}
\end{equation}
so that the total change in proton energy per particle is
\begin{equation}
d u_i = -d e_{w,i\perp}-d e_{w,i\parallel}\equiv - d e_{w,i\,\rm tot}~,\label{eq:u_ion}
\end{equation}
which simply states that the energy lost by protons is transferred to proton waves. Following \sect{phys}, the change in specific proton entropy is
\begin{eqnarray}
d s_i & = \left(\frac{1}{2}d\ln T_{i,\parallel}\right)\left(1-\frac{T_{i,\parallel}}{T_{i,\perp}}\right)-\frac{de_{w,i \,{\rm tot}}}{T_{i,\perp}}\,\label{eq:prot1} \\
d s_i & = \left(-d\ln T_{i,\perp}\right)\left(\frac{T_{i,\perp}}{T_{i,\parallel}}-1\right)-\frac{de_{w,i \,\rm{tot}}}{T_{i,\parallel}}\,\label{eq:prot2}
\end{eqnarray}
where the two expressions are equivalent, as with \eq{dselec} and \eq{dsperp}. We now need to specify $d e_{w,i\,\rm tot}$, i.e.,  the energy per proton {\it transferred} to proton modes. As we anticipated in \sect{phys}, this is not equal to the energy {\it residing} in proton waves, since some fraction of that is being used to perform work on the electrons. In the remainder of this section, we denote as $n$ and $B$ the density and magnetic field fluctuations induced by proton waves. Since the scale of the perturbations is much larger than the electron gyroradius, the fluctuations perform work on the electrons, so that \eq{worke} for electrons becomes
\be
\!\!\!\!\!\!dw_e=T_{e,\perp}d\ln B+T_{e,\parallel}d\ln\left(\frac{n}{B}\right)\equiv d{w_{e,\perp}}+d{w_{e,\parallel}}~.
\ee
This energy increase in the electrons is at the expense of the energy in proton waves, so that the residual energy per particle residing in proton waves will be
\be
d e_{w,i}=d e_{w,i\,\rm tot}-d{w_{e,\perp}}-d{w_{e,\parallel}}~,\label{eq:u_ionwave}
\ee
and the energy equation for protons reads
\be\label{eq:proten}
du_{i}=-d{w_{e,\perp}}-d{w_{e,\parallel}}-d e_{w,i}\,
\ee
where the three terms on the right hand side can be explicitly measured in our simulations. 

\fig{undriven1Dplot}(e) and (g) demonstrate that our heating model works remarkably well for protons (later on, we will show that it also works for electrons). In \fig{undriven1Dplot}(e), the blue solid line is the proton entropy change measured directly from the simulation, using the distribution function as we did in \eq{sdef}. It matches extremely well the prediction obtained by integrating the right hand side of the proton entropy equation, \eq{prot1} or \eq{prot2} (see the orange dotted line in \fig{undriven1Dplot}(e)). The agreement is also remarkable as regard to the proton energy equation, \eq{proten}. In \fig{undriven1Dplot}(g), the proton energy loss $-\Delta u_i=-\int du_i$ is indicated as a green line. From \eq{proten}, this should be equal to $\Delta e_{w,i}+w_\perp+w_\parallel$, where we have defined $\Delta e_{w,i}=\int de_{w,i}$, $w_\perp=\int d w_{e,\perp}$ and $w_\parallel=\int d w_{e,\parallel}$. In fact, the green line nearly overlaps with the red curve.


The growth of  proton cyclotron waves provides a source of field amplification and density perturbations that can perform work on the electrons. Indeed,
\fig{undriven1Dplot}(h) shows that during the exponential phase of the proton cyclotron instability ($4\lesssim \Omega_{ci}t\lesssim  7$), the proton waves increase the electron perpendicular energy (i.e., $dw_{e,\perp}>0$, see the blue line in Figure \ref{fig:undriven1Dplot}(h)) and decrease the parallel component (i.e., $dw_{e,\parallel}<0$, see the orange line in Figure \ref{fig:undriven1Dplot}(h)). This leads to a temperature anisotropy $T_\perpe>T_\pare$ (compare the blue and orange lines in Figure \ref{fig:undriven1Dplot}(d) at $\Omega_{ci}t\sim5$), which can be equivalently explained
as a result of the conservation of the first and second adiabatic invariants in the growing fields of the proton cyclotron waves. The resulting electron anisotropy 
is sufficiently strong to trigger the growth of whistler waves.

While the presence of whistler waves is hard to identify by eye in the space-time diagrams of Figure \ref{fig:undriven2Dplot}(b) and (c), due to the dominance of  proton cyclotron modes, we can extract their signature by applying a filter in frequency and wavenumber, as done in \sect{whistler}.
Figure \ref{fig:undriven2Dplot}(e) and (f) show the power spectra of $\delta B_x$ and $ \delta B_z$. Most of the power is concentrated near the origin at low frequencies and long wavelengths, associated with the proton cyclotron mode. However, we can still identify a significant peak  around $\omega \simeq 13\, \Omega_{ci}\simeq 0.3\,\Omega_{ce}$ and $k_y \simeq 0.3 \,\omega_{pe}/c$. In analogy with the discussion in \sect{whistler}, we associate this peak  with electron whistler waves. When applying a high-pass filter whose frequency and wavelength cuts are shown as red dashed lines in Figure \ref{fig:undriven2Dplot}(e) and (f), we recover in the space-time diagrams of Figure \ref{fig:undriven2Dplot}(h) and (i) the typical spatial and temporal patterns of electron whistler waves. As expected, most of the electron whistler activity takes place near the end of the exponential growth of proton waves, at $5\lesssim \Omega_{ci}t\lesssim 7$ (see also the temporal evolution of the  energy in whistler waves in \fig{undriven1Dplot}(b)). In this time interval, the electron anisotropy exceeds the threshold of whistler instability in the whole simulation domain (\fig{undriven2Dplot}(g)).

This period also corresponds to a rapid increase of the electron specific entropy, as measured directly from the electron distribution function (blue solid line in Figure \ref{fig:undriven1Dplot}(f)). This is expected, since whistler waves provide the pitch-angle scattering required to break adiabatic invariance, which (together with the sustained electron anisotropy, see \fig{undriven1Dplot}(d) and \fig{undriven2Dplot}(d) at $5\lesssim \Omega_{ci}t\lesssim 7$) drives efficient entropy generation. Based on our model in \sect{phys}, the electron specific entropy should increase as
\be\label{eq:dseions}
ds_e 
& = &\left[\frac{1}{2}d\ln \left(\frac{T_{e,\parallel}}{(n/B)^2}\right)\right]\left(1-\frac{T_{e,\parallel}}{T_{\perpe}}\right)-\frac{d e_{w,e}}{T_{\perpe}}~,
\ee
which follows from \eq{dselec} (an equivalent form can be obtained from \eq{dsperp}). Here, we have used the condition $d e_{w,e \,\rm  tot}=d e_{w,e}$ for electrons. The comparison of the measured entropy increase (blue solid line in \fig{undriven1Dplot}(f)) with the predicted entropy change (orange dotted line in \fig{undriven1Dplot}(f)) provides another validation of our heating model. 

While most of the electron entropy production happens near the end of the exponential growth of proton waves, a moderate increase of the electron entropy also occurs during the secular stage (i.e., at $\Omega_{ci}t\gtrsim 10$). Here, the oscillating cyclotron fluctuations are sloshing electrons around and can occasionally excite local patches of electron anisotropy that exceed the whistler threshold (e.g., see Figure \ref{fig:undriven2Dplot}(g) at $\Omega_{ci}t\simeq 11$ and $y\simeq 175 \, \compe$). In the same region, we can identify a short episode of electron whistler activity, particularly in $\delta  B_z$ in \fig{undriven2Dplot}(i). These sporadic episodes of anisotropy-driven whistler waves further increase the electron entropy.

It is worth noticing, though, that at late times the box-averaged electron anisotropy switches sign, with $T_\pare\gtrsim T_\perpe$ (in \fig{undriven1Dplot}(d), at $\Omega_{ci}t\gtrsim 9$), so the opportunities for whistler growth are fewer. This behavior is consistent with the conservation of the first and second adiabatic invariants in the decaying field of the proton cyclotron waves, leading to an increase in $T_\pare$ and a decrease in $T_\perpe$ (as indeed seen in \fig{undriven1Dplot}(d) at late times). The same ``inverted'' anisotropy with $T_\pare\gtrsim T_\perpe$ is seen in the far downstream of our reference shock run (\fig{elecs}(f)), accompanying the decay of the proton modes. We remark that electron entropy production is still possible when $T_\pare\gtrsim T_\perpe$, as long as the anisotropy is large enough to exceed the threshold of the firehose instability, which would then provide the mechanism for breaking adiabatic invariance. We have verified with expanding box simulations similar to the ones reported in \sect{ramp} (not shown) that once the system exceeds the firehose threshold, the electron entropy rapidly increases.

Finally, by measuring directly the energy in whistler waves, we can also validate the electron energy equation
\be\label{eq:dueions}
du_{e}=d{w_{e,\perp}}+d{w_{e,\parallel}}-d e_{w,e}~.
\ee
Once again, the time-integral of the left hand side matches very well the time-integral of the right hand side (compare green and red curves in \fig{undriven1Dplot}(h)), i.e., the change of electron internal energy can be accounted for by the total work done by the proton waves and the energy lost to generate electron whistler waves. 

In summary, we have demonstrated that efficient electron entropy production can occur even in the absence of a large-scale compression. Magnetic and density fluctuations sourced by anisotropic protons drive electrons to become anisotropic, with $T_\perpe>T_\pare$. The electron anisotropy is relaxed by the growth of whistler waves, which break the electron adiabatic invariance and mediate the production of electron entropy. In the next subsection, we show that the resulting electron entropy increase is nearly independent of the proton-to-electron mass ratio.

\begin{figure}[tbh]
\begin{center}
\includegraphics[width=0.36\textwidth]{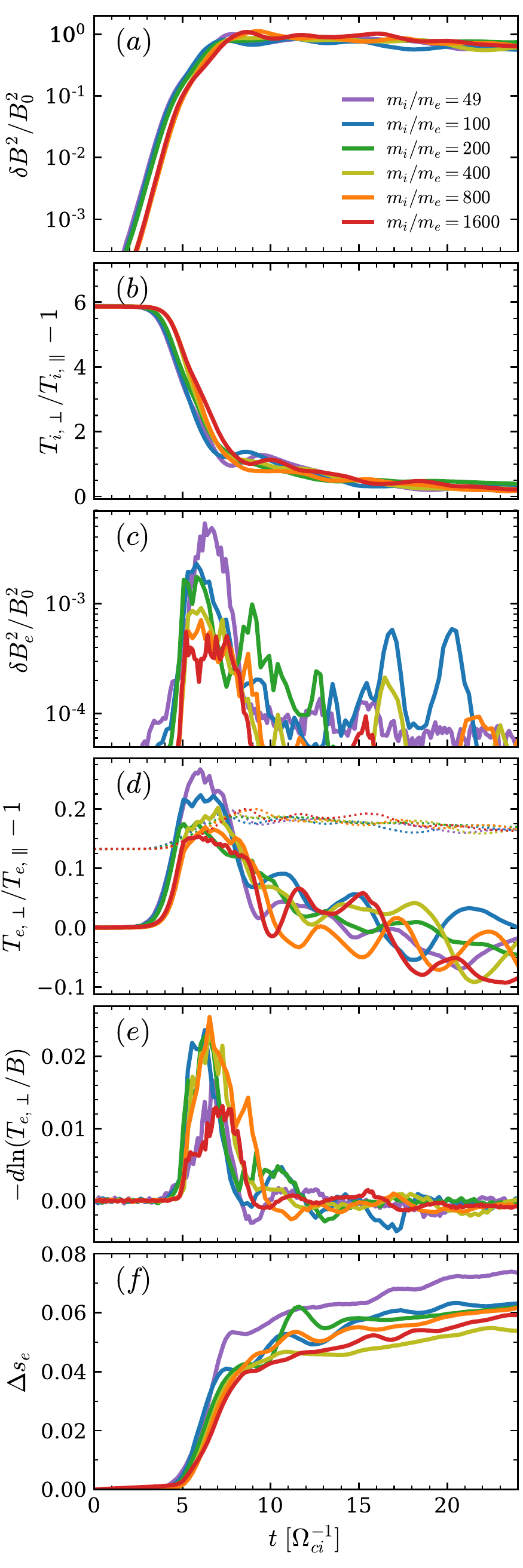}
\end{center}
\caption{Dependence on mass ratio (up to $m_i/m_e=1600$) of various space-averaged quantities in a 1D periodic box with anisotropic protons (the legend is in panel (a)). The background field is aligned with the box (in-plane configuration). We plot: (a) energy in magnetic field fluctuations, normalized to the energy of the initial  field; (b) proton temperature anisotropy; (c) energy in electron-scale field fluctuations; (d) electron temperature anisotropy (solid lines) and threshold condition for the electron whistler instability (dotted lines with the same color coding as the solid lines); (e) rate of violation of the electron adiabatic invariance $-d\ln (T_\perpe/B)$; (f) electron entropy change, measured from the electron distribution function.}\label{fig:undriven_massratio}
\end{figure}

\subsection{Dependence on the Mass Ratio}\label{sec:mass2}
In this subsection, we explore with periodic boxes initialized with anisotropic protons how the development of the proton cyclotron instability can lead to electron irreversible heating. We vary the mass ratio from 49 up to 1600, as indicated in the legend of \fig{undriven_massratio}(a). Since the fastest growing mode of the proton cyclotron instability has wavevector $\sim \omega_{pi}/c$, we increase the number of cells in our computational domain as $\propto \sqrt{m_i/m_e}$, to include the same number of proton skin depths for all values of $m_i/m_e$ (and so, the same number of proton cyclotron wavelengths).

Figure \ref{fig:undriven_massratio} compares the runs. Panel (a) shows the time evolution of the wave magnetic energy, which is dominated by proton cyclotron modes. Panel (b) shows the evolution of the proton temperature anisotropy, which reduces strongly at $\Omega_{ci}t\gtrsim 4$ by pitch angle scattering off the strong cyclotron waves. Unsurprisingly, since these two quantities are related to protons, their evolution is almost identical for different mass ratios. As long as the mass ratio is sufficiently large to adequately separate electron and proton scales, the proton cyclotron instability --- whose polarization is resonant with protons, but non-resonant with electrons --- is not affected by electron physics.

As in the case $m_i/m_e=49$ discussed above, the growth of the proton cyclotron instability  induces an electron temperature anisotropy  with $T_\perpe>T_\pare$ (Figure \ref{fig:undriven_massratio}(d)) which excites electron whistler waves (Figure \ref{fig:undriven_massratio}(c)),\footnote{We isolate the magnetic energy associated with whistler waves by applying a high-pass filter for frequency higher than $0.5 \Omega_{ce}$ and wavelength shorter than $35 c/\omega_{pe}$. } and facilitate electron entropy increase  (Figure \ref{fig:undriven_massratio}(f)) by violating the electron adiabatic invariance (Figure \ref{fig:undriven_massratio}(e)). The dependence of the peak electron anisotropy on mass ratio for $m_i/m_e\lesssim 200$ can be understood from the same argument we have presented in \sect{ramp}: in units of the electron gyration period, the growth of proton waves (or the compressed field, for \sect{ramp}) is faster at lower $m_i/m_e$, which leads to an overshoot in electron anisotropy beyond the threshold of whistler marginal stability. The overshoot is more pronounced for lower $m_i/m_e$. Due to the higher electron anisotropy, more free energy is available for the growth of whistler waves at lower $m_i/m_e$ (see the trend from purple to green in \fig{undriven_massratio}(c) at $\Omega_{ci}t\sim 6$). Because of the higher electron anisotropy and stronger whistler waves, the electron entropy increases slightly more at lower mass ratios (in particular, see the purple line in \fig{undriven_massratio}(f) for $m_i/m_e=49$).

On the other hand, the electron physics shows no appreciable dependence on mass ratio for $m_i/m_e\gtrsim 400$. The peak electron anisotropy at $5\lesssim\Omega_{ci}t\sim 7$ saturates at the threshold of whistler marginal stability (indicated in \fig{undriven_massratio}(c) by the dotted lines, with the same color coding as the solid lines). As a consequence, the peak strength of whistler waves is nearly independent of mass ratio (see \fig{undriven_massratio}(b) in the same time interval), and the resulting entropy increase is the same for all mass ratios $m_i/m_e\gtrsim 400$ (\fig{undriven_massratio}(f)). Even for $m_i/m_e=49$, the electron entropy increase at the final time is only $\sim30\%$ higher than for $m_i/m_e=1600$.

\begin{figure}[tbh]
\begin{center}
\includegraphics[width=0.5\textwidth]{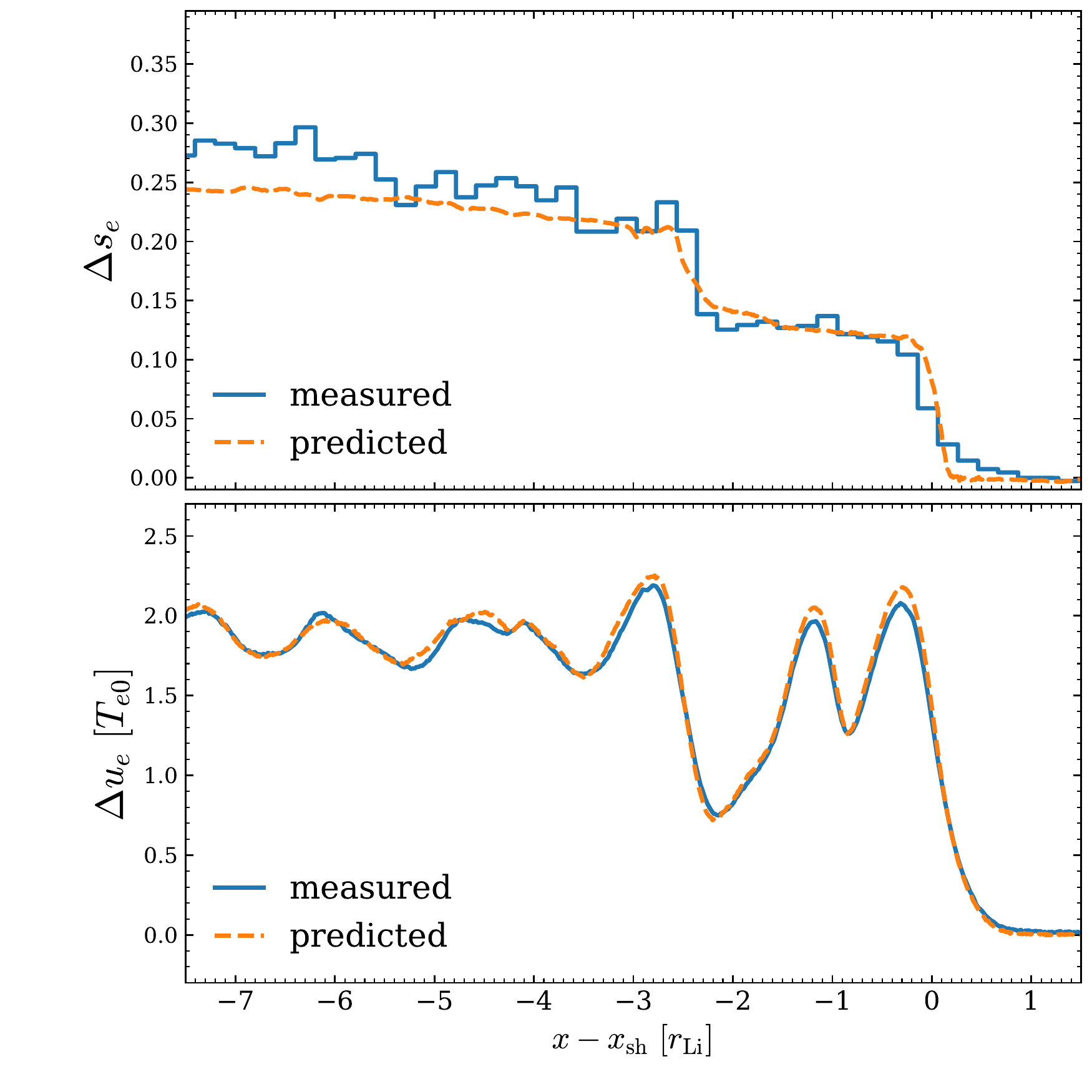}
\end{center}
\caption{Validation of the heating model in our reference shock simulation at $\Omega_{ci}t=25.6$. In the top panel, we compare the $y$-averaged  electron entropy change measured with the electron distribution function as in \eq{sdef} (blue solid line) with the predicted change based on \eq{dsperp} (orange dashed line). The differential terms on the right hand side of \eq{dsperp} are calculated from the difference of neighboring cells along the $x$ direction. In the bottom panel, we compare the $y$-averaged  electron energy change measured directly from our simulation (blue solid line) with the predicted increase based on \eq{energyconserv} (orange dashed line). For both entropy and internal energy, the agreement between the model and the simulation results is remarkably good.}\label{fig:check_in_shock}
\end{figure}

\begin{figure}[tbh]
\begin{center}
\includegraphics[width=0.5\textwidth]{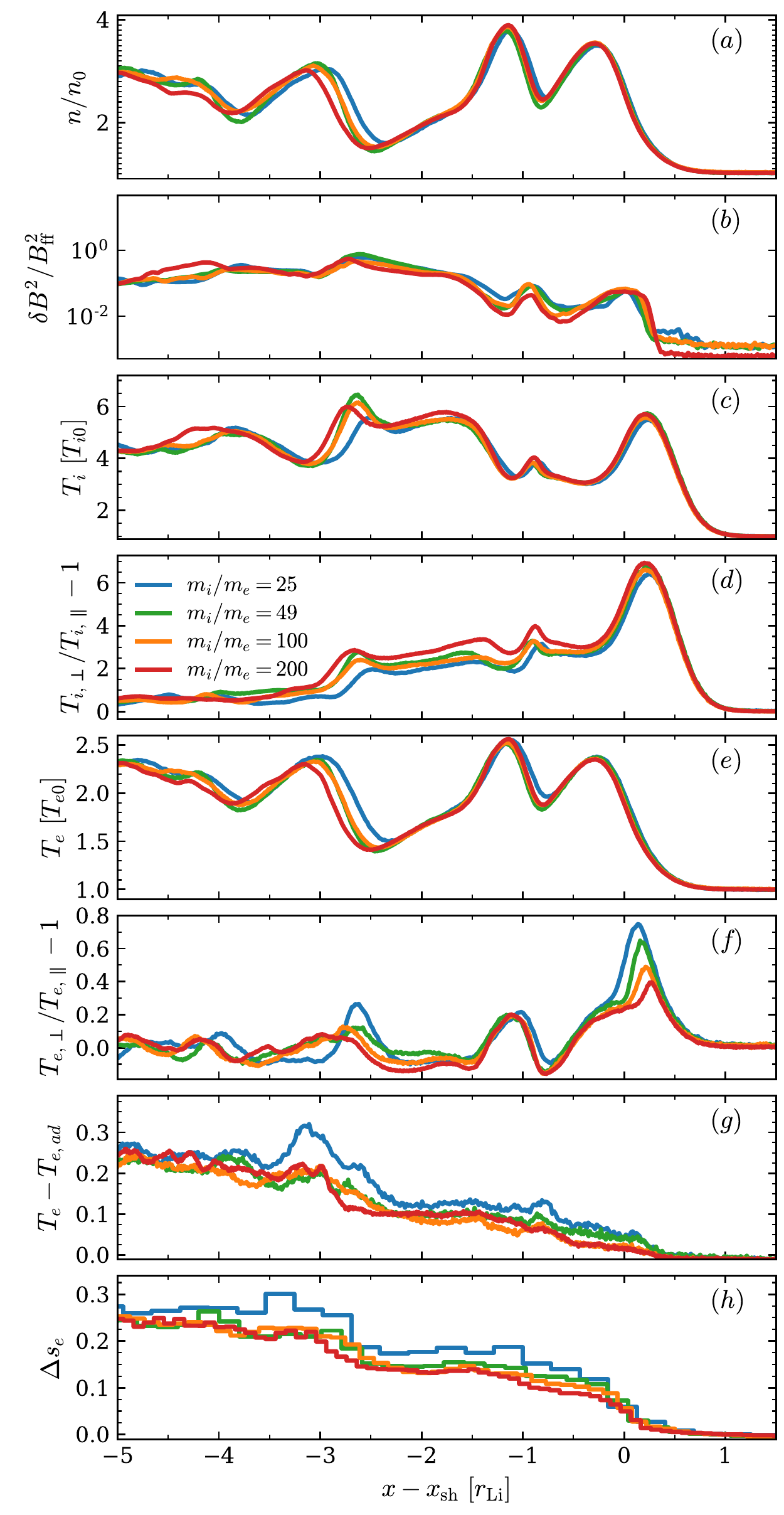}
\end{center}
\caption{Dependence on mass ratio (up to $m_i/m_e=200$) of shock simulations at $t=13.1\,\Omega_{ci}^{-1}$ (the legend is in panel (d)). Along the shock direction of propagation, we plot the $y$-averaged profiles of: (a) number density; (b) energy in magnetic fluctuations, normalized to the energy of the frozen-in  field; (c) mean proton temperature; (d) proton temperature anisotropy; (e) mean electron temperature; (f) electron temperature anisotropy; (g) excess of electron temperature beyond the adiabatic prediction for an isotropic gas; (h) change in electron entropy. The increase in electron entropy is nearly insensitive to the mass ratio.
}
\label{fig:checkmass}
\end{figure}

\section{Validation of the Electron Heating Physics in Shocks}\label{sec:back}
We are now in a position to validate our heating model in full shock simulations. In \sect{ramp} and \sect{waves}, we have demonstrated that our heating model provides an excellent description of the change in electron energy and entropy for two physical scenarios: if electrons are subject to a large-scale compression, as in the shock ramp; and if electrons are driven to temperature anisotropy by the growth of proton-driven modes, as observed in the far downstream. Since the two scenarios correspond to the two locations where the entropy profile in shocks shows the fastest  increase, we expect that our model will properly capture the electron heating physics in our reference shock run described in \sect{shock}.

In the top panel  of \fig{check_in_shock}, we compare the electron entropy profile measured directly from the phase space distribution function as in \eq{sdef} (solid blue line), with the entropy change predicted by \eq{dsperp} (dashed orange line). The differential terms on the right hand side of \eq{dsperp} are calculated from the difference of neighboring cells along the $x$ direction. The agreement between the measured entropy profile and the predicted one is remarkably good (with the exception of the far downstream region, where numerical heating of electrons might be responsible for the discrepancy, see \app{ppc}). In particular, the theory correctly predicts the location and magnitude of the two sites of fastest entropy growth: in the shock ramp, where electron irreversible heating is induced by the shock-compression of density and magnetic field (in analogy to the scenario we have studied in \sect{ramp}); and at a distance of $\sim 2.5\rli$ behind the shock, where a large-scale density and field compression co-exists with the growth of proton-driven waves, the latter contributing to further magnetic field amplification. The agreement of theory and measurement at this location is then a {\it combined} validation of the two scenarios described in \sect{ramp} and \sect{waves}, confirming that our model holds regardless of what drives the field amplification (and so, the resulting electron anisotropy).

In addition, in the bottom panel of \fig{check_in_shock} we show that the change in electron energy per particle (blue solid line) is predicted extremely well by our heating model (orange dashed line, following \eq{energyconserv}).

\subsection{Dependence on the Mass Ratio}
In the periodic box runs of \sect{ramp} and \sect{waves}, we have extended our study to realistic mass ratios, showing that the entropy increase at $m_i/m_e=1600$ is only $\sim 30\%$ smaller than for the choice $m_i/m_e=49$ of our reference shock simulation. In \fig{checkmass}, we investigate the dependence of the electron physics in our full shock simulations on the mass ratio, from $m_i/m_e=25$ up to $m_i/m_e=200$ (as indicated in the legend of panel (d)). We typically employ 32 computational particles per cell, with the exception of $m_i/m_e=200$, where we use 64 particles per cell to keep numerical heating under control. We keep the upstream electron temperature fixed at $k_{\rm B}T_{e0}=10^{-2} m_e c^2$, so that electrons stay safely non-relativistic. This implies that the plasma inflow velocity is slower with increasing mass ratio, as $\propto \sqrt{m_e/m_i}$. 

The proton physics is expected to be the same regardless of mass ratio, and in fact the profiles of density (panel (a)), proton temperature (panel (c)) and proton anisotropy (panel (d)) are nearly the same for all mass ratios. The same holds for the wave magnetic energy at $\xshnorm\lesssim -0.5$  (panel (b)), where proton-driven modes dominate (see \sect{shock} for details).

On the other hand, the peak electron anisotropy at the shock (panel (e)) is systematically lower for higher mass ratios, in perfect agreement with the trend observed in the periodic box experiments of \fig{compressbox_massratio}. Despite the pronounced difference in peak anisotropy, \fig{compressbox_massratio}(d) showed that the entropy increase until $t\sim \Omega_{ci}^{-1}$ was only marginally lower at higher $m_i/m_e$. This trend (and the weak mass ratio dependence) is confirmed by the profiles of electron entropy in the shock ramp shown in \fig{checkmass}(h). Overall, \fig{checkmass}(h) confirms the results of our periodic box experiments, namely, the electron entropy increase is nearly independent of mass ratio (with the exception of the lowest mass ratio $m_i/m_e=25$ presented in \fig{checkmass}). Even though our shock simulations only extend up to $m_i/m_e=200$, the results of our periodic runs in \sect{ramp} and \sect{waves} suggest that the same conclusion should hold up to the realistic mass ratio.

\section{Summary and Discussion}\label{sec:disc}
In this work, we have investigated by means of analytical theory and 2D PIC simulations the electron heating physics in perpendicular low Mach number shocks, in application to merger shocks in galaxy clusters. While most of the electron heating is adiabatic --- induced by shock-compression of the upstream magnetic field --- we direct our attention to the electron entropy increase, i.e., to the production of irreversible electron heating. 

We find that, in analogy to the so-called ``magnetic pumping'' mechanism, two basic ingredients are needed for electron irreversible heating: (\textit{i}) the presence of a temperature anisotropy, induced by field amplification coupled to adiabatic invariance; and (\textit{ii}) a mechanism to break the adiabatic invariance itself. 

We have demonstrated that, in our reference shock with sonic Mach number $M_s=3$ and plasma beta $\beta_{p0}=16$, efficient electron entropy production occurs at two major sites: at the shock ramp, where density compression coupled to flux freezing leads to field amplification and a high degree of electron anisotropy; and farther downstream, where density compression and long-wavelength magnetic waves induced by the proton temperature anisotropy are both contributing to magnetic field growth. Regardless of the origin of field amplification, electrons are driven to a large degree of temperature anisotropy, exceeding the threshold of the electron whistler instability. The resulting growth of electron whistler waves --- whose presence is one of the common denominators of the two sites mentioned above --- causes the violation of the electron adiabatic invariance, and allows for efficient entropy production.

Our model is in excellent agreement with the measured electron entropy increase, which can be quantified directly from the electron distribution function in our simulations. The agreement holds for our reference shock simulation, as well as for controlled periodic box experiments meant to reproduce the shock conditions at the two major sites of entropy production. In particular, the shock physics in the ramp can be replicated in a periodic box where the PIC equations are modified to allow for a continuous large-scale compression, as in \citet{Sironi2015,Sironi2015a}. Also, the physics of anisotropy-driven proton waves, and the resulting electron irreversible heating, can be conveniently studied in a periodic box initialized with anisotropic protons, with a degree of anisotropy inspired by the shock simulation. The advantage of the periodic domains is twofold: (\textit{i}) they allow for a more direct control of the relevant physics; (\textit{i}) and, due to less demanding computational requirements, they permit to extend our investigation up to the realistic mass ratio. We have then be able to ascertain that the entropy increase has only a weak dependence on mass ratio (less than $\sim 30\%$ decrease, as we increase the mass ratio from $m_i/m_e=49$ up to $m_i/m_e=1600$).

Finally, we remind that in this paper (the first of a series), we have only focused on one representative set of shock parameters, fixing the Mach number $M_s=3$ and the plasma beta $\beta_{p0}=16$. In a forthcoming work (X. Guo et al., in preparation) we will explore the dependence of our conclusions on sonic Mach number and plasma beta, and we will discuss the implications of our results for observations of galaxy cluster shocks.

\acknowledgements
This work is supported in part by the Black Hole Initiative at Harvard University though a grant from the Templeton Foundation.  XG and RN acknowledge support from NASA TCAN NNX14AB47G and NSF grant AST 1312651. LS acknowledges
support from DoE DE-SC0016542, NASA
Fermi NNX16AR75G, NSF ACI-1657507, and NSF AST-1716567. The simulations were performed on Habanero at Columbia, the BHI cluster at the Black Hole Initiative, NASA High-End Computing (HEC) Program through the NASA Advanced Supercomputing (NAS) Division at Ames Research Center, and NSF XSEDE resources (grant TG-AST080026N). 

\appendix

\begin{figure}[tbh]
\begin{center}
\includegraphics[width=0.5\textwidth]{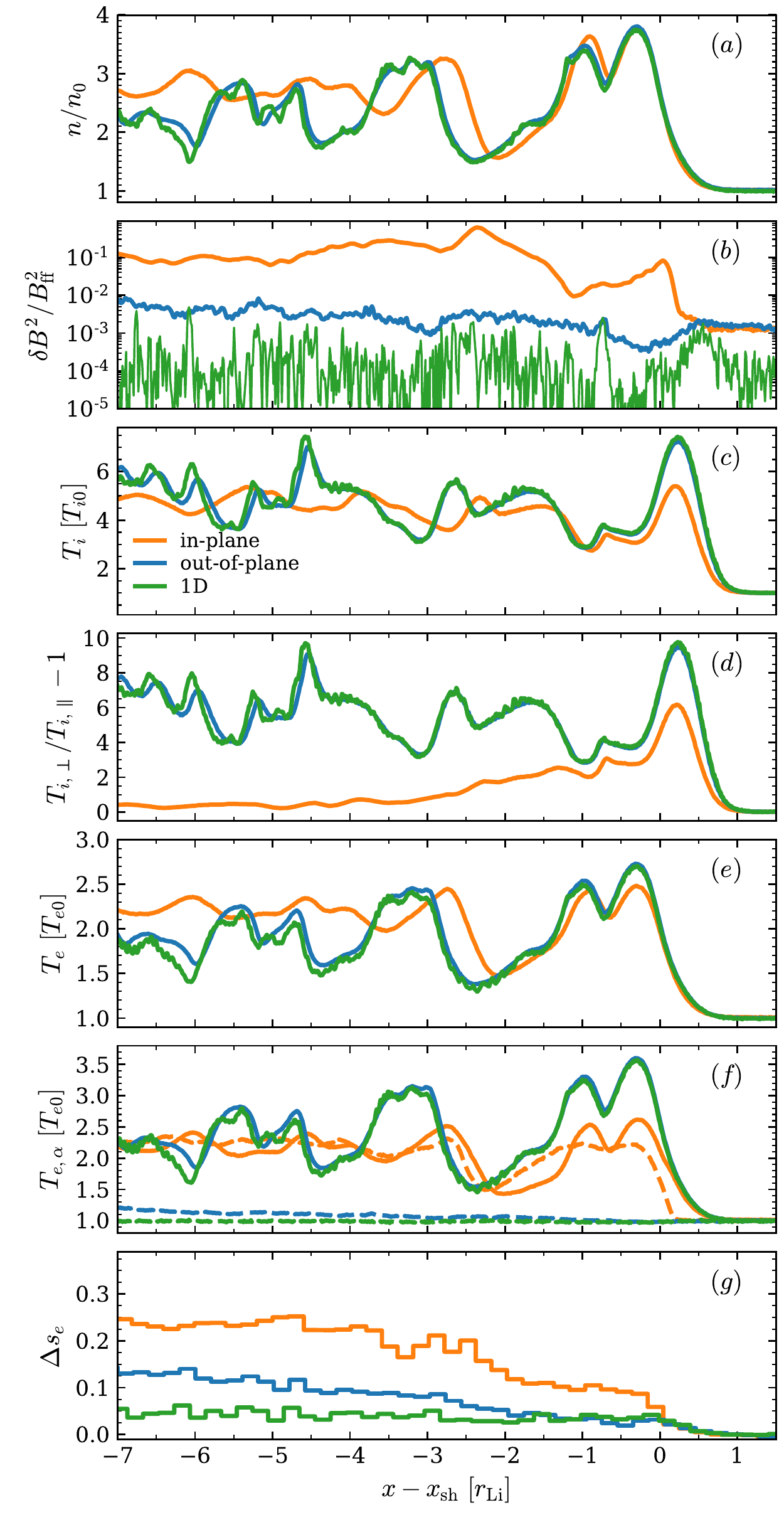}
\end{center}
\caption{Comparison at $\Omega_{ci}t=23.1$ between two 2D simulations with in-plane (orange) or out-of-plane (blue) fields and a 1D simulation (green), as indicated in the legend of panel (c). Along the shock direction of propagation, we plot the $y$-averaged profiles of: (a) number density; (b) energy in magnetic fluctuations, normalized to the energy of the frozen-in  field; (c) mean proton temperature; (d) proton temperature anisotropy; (e) mean electron temperature; (f) electron temperature perpendicular (solid) and parallel (dashed) to the bakground field; (g) change in electron entropy.}\label{fig:shock_inoutplane}
\end{figure}
\section{Comparison between in-plane and out-of-plane magnetic field geometries}\label{sec:outplane}
In the 2D shock simulations presented in the main body of the paper, we have initialized the upstream field in the $xy$ plane of the simulation (``in-plane'' geometry). As we have discussed, this is instrumental in capturing the dominant wavevector of both proton and electron waves: the fastest growing mode of the proton cyclotron instability is aligned with the background field, and mirror modes are also naturally resolved if the magnetic field lies in the simulation plane; similarly, the dominant mode of the electron whistler instability is nearly parallel to the background field.

Given that the heating mechanism that we propose rely on such waves for breaking the electron adiabatic invariance (in the case of whistler waves) or for amplifying the magnetic field, thus leading to irreversible electron heating (in the case of proton modes), we expect that the alternative ``out-of-plane'' geometry, in which the field is initialized along the $z$ direction, will lead to weaker electron heating. This is confirmed by \fig{shock_inoutplane}: there, orange lines refer to our reference 2D simulation with in-plane fields, blue lines to a 2D simulation with out-of-plane fields, and green lines to a 1D simulation. The physical and numerical parameters of the two 2D runs are the same as in our reference run (of course, apart from the field orientation). The 1D simulation has the same physical parameters, but a higher number of particles per cell (5000 per species).

As expected, the 2D out-of-plane case is remarkably similar to 1D results (compare blue and green lines). In both cases, both protons and electrons stay highly anisotropic (panels (d) and (f)), due to the lack of anisotropy-driven waves (and in fact, the wave energy in panel (b) does not appreciably exceed noise levels). This should be contrasted with the in-plane case (orange lines), where both electron and proton anisotropies get reduced by the effect of strong self-generated waves. As a consequence, the entropy increase in the in-plane case (orange line in panel (g)) is much more pronounced than in the out-of-plane run (blue), which in turn is quite similar to the 1D result (green).\footnote{The deviation of the blue and green lines in the far downstream region of panel (g) is likely to come from numerical noise in the 2D out-of-plane simulation.}

\begin{figure}[tbh]
\begin{center}
\includegraphics[width=0.5\textwidth]{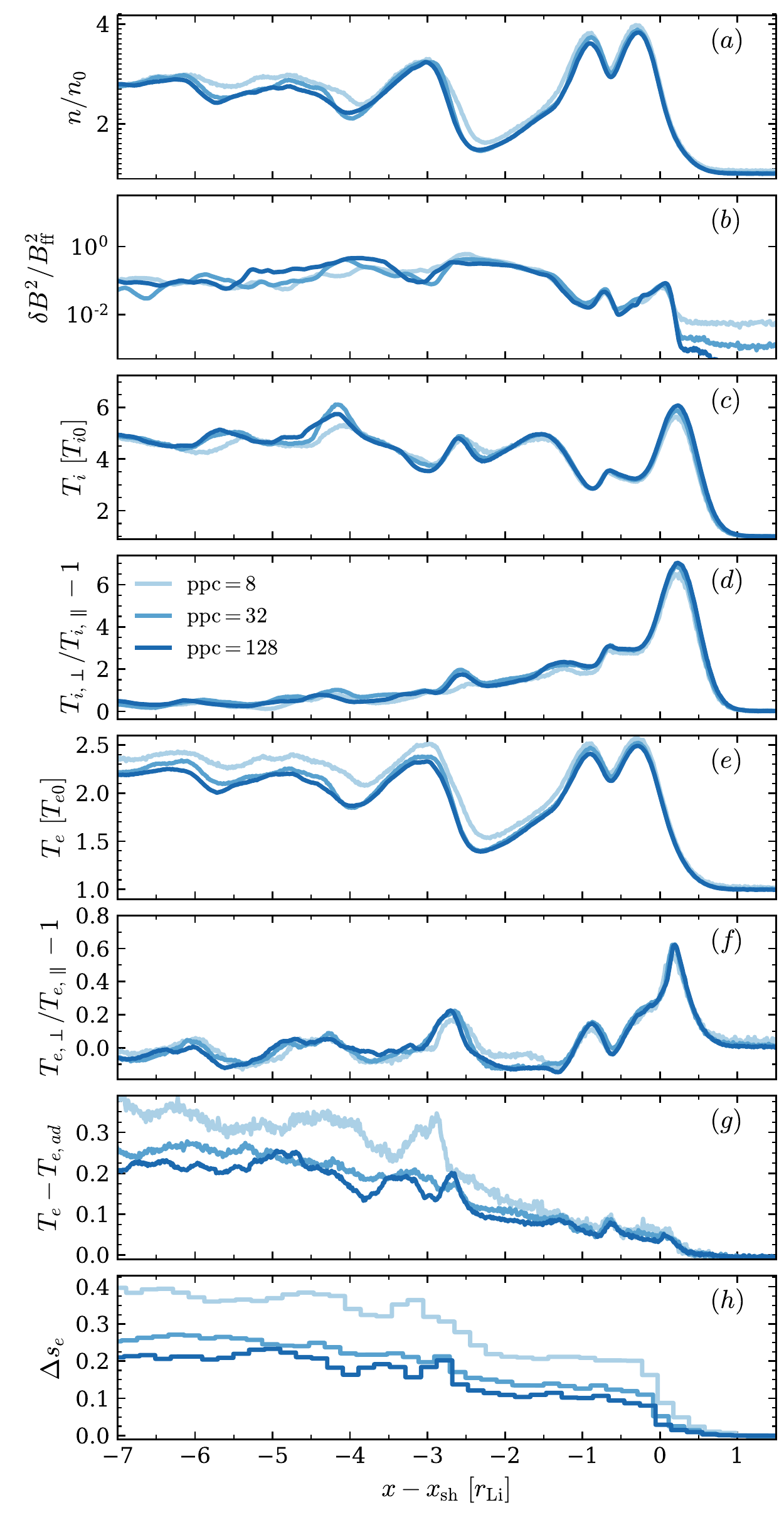}
\end{center}
\caption{Comparison at $\Omega_{ci}t=15.8$ of three runs with the same physical parameters (as in our reference shock run) but a different number of particles per cell, as indicated in the legend of panel (d). Along the shock direction of propagation, we plot the $y$-averaged profiles of: (a) number density; (b) energy in magnetic fluctuations, normalized to the energy of the frozen-in  field; (c) mean proton temperature; (d) proton temperature anisotropy; (e) mean electron temperature; (f) electron temperature anisotropy; (g) excess of electron temperature beyond the adiabatic prediction for an isotropic gas; (h) change in electron entropy.}\label{fig:shock_ppc}
\end{figure}
\section{Dependence on the number of computational particles per cell}\label{sec:ppc}
In a two-temperature plasma, with protons hotter than electrons, numerical noise will tend to heat the electrons, even in the absence of any physical effect. It is therefore important to check that our results are converged with respect to the number of computational particles per cell, whose value controls the noise level of PIC simulations, and so the rate of numerical electron heating. In \fig{shock_ppc}, we compare our results for three choices of the number of particles per cell (including both species), from 8 (light blue) to 128 (dark blue), as shown in the legend of panel (d). \fig{shock_ppc} shows that the proton physics is largely independent from the number of particles per cell (panels (a), (c) and (d)). On the other hand,
panel (b) shows that for 8 particles per cell the noise level of field fluctuations is not negligible, as compared to the physical fields (see the light blue line in panel (b) ahead of the shock). As  a result, electrons are heated due to numerical artifacts (light blue line in panels (g) and (h)), to a temperature much larger than in runs with a higher number of particles per cell. The comparison of the runs with 32 and 128 particles per cell shows solid evidence of convergence, even in the profiles of irreversible electron heating (panels (g) and (h)), which are most sensitive to numerical noise. Still, a small  difference persists between the entropy profiles obtained with 32 and 128 particles per cell (compare the medium-blue with the dark-blue line in panel (h)). We argue that the deviation of our model from the measured entropy profile in the top panel of \fig{check_in_shock} might be largely explained by numerical effects acting far behind the shock.

\bibliographystyle{apj}
\bibliography{Mendeley,heating}
\end{document}